\documentclass{article}

\usepackage{arxiv}
\usepackage{amsmath}
\usepackage[utf8]{inputenc} 
\usepackage[T1]{fontenc}    
\usepackage{hyperref}       
\usepackage{url}            
\usepackage{booktabs}       
\usepackage{amsfonts}       
\usepackage{nicefrac}       
\usepackage{microtype}      
\usepackage{lipsum}
\usepackage{graphicx}
\graphicspath{ {./images/} }

\usepackage{lineno}
\usepackage{siunitx}
\usepackage{url}
\usepackage{cite}
\usepackage{hyperref}

\title{Separating edges from microstructure in X-ray dark-field imaging: Evolving and devolving perspectives via the X-ray Fokker--Planck equation}

\author{
 Samantha J. Alloo \\
 School of Physics and Astronomy, Monash University, Victoria, Australia \\
 School of Physical and Chemical Sciences, University of Canterbury, Christchurch, New Zealand\\
 \texttt{samantha.alloo@monash.edu} \\
   \And
 David M. Paganin \\
 School of Physics and Astronomy, Monash University, Victoria, Australia \\
  \And
 Michelle K. Croughan \\
 School of Physics and Astronomy, Monash University, Victoria, Australia \\
   \And
 Jannis N. Ahlers \\
 School of Physics and Astronomy, Monash University, Victoria, Australia \\
    \And
 Konstantin M. Pavlov \\
 School of Physical and Chemical Sciences, University of Canterbury, Christchurch, New Zealand\\
 School of Physics and Astronomy, Monash University, Victoria, Australia \\
 School of Science and Technology, University of New England, Armidale, Australia \\
 \And
 Kaye S. Morgan \\
 School of Physics and Astronomy, Monash University, Victoria, Australia \\
}

\begin{document}
\maketitle
\begin{abstract} 
A key contribution to X-ray dark-field (XDF) contrast is the diffusion of X-rays by sample structures smaller than the imaging system’s spatial resolution; this is related to position-dependent small-angle X-ray scattering. However, some experimental XDF techniques have reported that XDF contrast is also generated by resolvable sample edges. Speckle-based X-ray imaging (SBXI) extracts XDF by analyzing sample-imposed changes to a reference speckle pattern's visibility. We present an algorithm for SBXI (a variant of our previously developed Multimodal Intrinsic Speckle-Tracking (MIST) algorithm) capable of separating these two physically different XDF contrast mechanisms. The algorithm uses what we call the \textit{devolving} Fokker--Planck equation for paraxial X-ray imaging as its forward model and then solves the associated multimodal inverse problem, to retrieve the attenuation, phase, and XDF properties of the sample. Previous MIST variants were based on the \textit{evolving} Fokker--Planck equation, which considers how a reference-speckle image is modified by the introduction of a sample. The devolving perspective instead considers how the image collected in the presence of the sample and the speckle membrane optically flows in reverse, to generate the reference-speckle image when the sample is removed from the system. We compare single- and multiple-exposure multimodal retrieval algorithms from the two Fokker--Planck perspectives. We demonstrate that the devolving perspective can distinguish between two physically different XDF contrast mechanisms, namely, unresolved microstructure- and sharp-edge-induced XDF. This was verified by applying the different retrieval algorithms to two experimental data sets -- one phantom sample and one organic sample. We anticipate that this work will be useful in (1) yielding a pair of complementary XDF images that separate sharp-edge diffuse scatter from diffuse scatter due to spatially random unresolved microstructure, (2) XDF computed tomography, where the strong edge XDF signal can lead to strong contaminating streaking artefacts, and (3) sample preparation, as samples will not need to be embedded since the strong XDF edge signal seen between the sample and air can be separated out.
\end{abstract}
\section{Introduction}
As hard X-rays travel through an object, micro- and macroscopic changes in the wavefield are generated. In the context of X-ray imaging, these changes can be described by X-ray attenuation, phase shift, and diffuse scatter \cite{paganin2006coherent}. Since the phase of an X-ray wavefield cannot be measured directly by the detector, special optics or experimental configurations are often required, leading to the development of phase-contrast X-ray imaging (PCXI) techniques \cite{paganin2019tutorials}. X-ray dark-field (XDF) imaging complements attenuation and phase shift images, as image contrast is generated through diffuse X-ray scatter. Conventionally, this X-ray diffusion has been attributed to position-dependent small-angle X-ray scattering (SAXS) from sample structures smaller than the spatial resolution, as well as structures that generate phase variations smaller than the spatial resolution (multiple refraction) \cite{wernick2003multiple, pfeiffer2008hard}. Additionally, it has been reported that sample edges can generate a retrievable XDF signal \cite{yashiro2010origin, yashiro2015effects}. XDF has proven applicability, for example, in food engineering \cite{nielsen2017quantitative,lim2022low,he2024nondestructive}, airport security \cite{miller2013phase}, and medicine \cite{stampanoni2011first,shimao2021x,aminzadeh2022imaging,gassert2023dark}. There are different experimental PCXI techniques capable of measuring XDF, for example, using free-space propagation \cite{gureyev2020dark,leatham2023x, ahlers2024x}, an analyzer crystal \cite{pagot2003method,wernick2003multiple,kitchen2010x}, grating-interferometry \cite{pfeiffer2008hard}, edge-illumination \cite{Endrizzi2014,endrizzi2015edge}, and techniques that introduce a single reference modulator, like two-dimensional (2D) period grid patterns \cite{how2022quantifying} or a random speckle membrane \cite{berujon2012x}. The definition and nature of the retrieved XDF signal varies between experimental techniques and can also depend on the retrieval algorithm used.

\begin{figure}[!b]
    \centering
    \includegraphics[width=\textwidth]{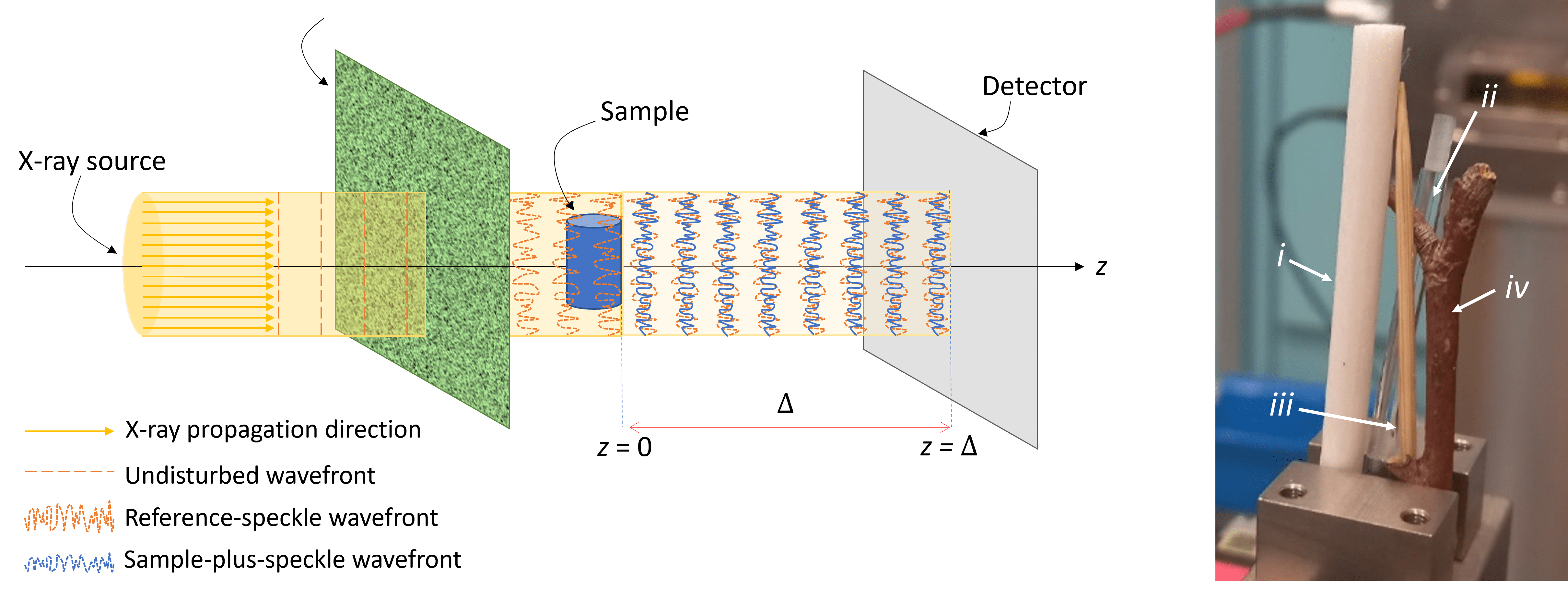}
    \caption{(left) Speckle-based X-ray imaging (SBXI) set-up for a monochromatic paraxial X-ray source. (right) The four-rod sample analyzed in this study, composed of \textit{i}) a reed diffuser stick, \textit{ii}) a polymethyl methacrylate (PMMA) rod, \textit{iii}) a toothpick, and \textit{iv}) a tree twig.
    }
    \label{fig:SBXI_Setup}
\end{figure}

Realized just over a decade ago in 2012 \cite{berujon2012two,morgan2012x}, speckle-based X-ray imaging (SBXI) is one of the newer XDF techniques. The experimental set-up for SBXI using a monochromatic paraxial X-ray source is shown in Fig.~\ref{fig:SBXI_Setup}. 
The technique followed closely from the 2D periodic grid approaches \cite{Wen2010,Bennett2010,Morgan2011,Morgan2013} after it was noticed the X-ray wavefield could be marked using a spatially random membrane instead of a periodic structure. This generalization alleviates the need to precisely fabricate periodic grids, which can be costly and limited in area. In SBXI, a spatially random membrane imprints the X-ray wavefield with resolvable intensity variations, known as speckles. In the presence of a sample, reference speckles are reduced in intensity, transversely shifted, and blurred after passing through the sample. These sample-induced speckle modifications can be tracked using a suitable algorithm to recover multimodal sample information. To retrieve information about sample-induced speckle modifications, two images are recorded: (1) a reference-speckle image, taken when only the random membrane is illuminated by the X-ray beam, and (2) a sample-plus-speckle image, taken when the sample is placed into the beam, distorting the reference speckle pattern.

As this study only considers SBXI using a paraxial monochromatic X-ray beam, which typically corresponds to a monochromated synchrotron source, the brief literature review below will focus on the relevant speckle-tracking algorithms. There are several speckle-tracking algorithms \cite{zdora2018state}, including X-ray speckle vector tracking (XSVT) \cite{berujon2012two}, mixed-XSVT \cite{berujon2016x,berujon2017near}, X-ray speckle scanning (XSS) \cite{wang2016high,wang2016synchrotron}, unified modulated pattern analysis (UMPA) \cite{zdora2017x}, and multimodal intrinsic speckle tracking (MIST) \cite{pavlov2020x}. Each method defines XDF slightly differently, but typically this is through a reduction in visibility, measured either by the standard deviation of the local intensity (XSVT) or stepping curve amplitude \cite{zdora2018state}. Therefore, the recovered XDF signal may differ depending on the exact multimodal algorithm used, the sensitivity of the SBXI experimental set-up, as well as the size of diffusing sample structures. This is demonstrated in Fig.~2 in Pavlov \textit{et al.} \cite{pavlov2020x}, where it is shown that the largest difference in the retrieved XDF signal using various speckle-tracking algorithms occurs for sample structures that are comparable in size to the spatial resolution of the imaging system where a significant portion of the signal is introduced by edges.

MIST defines the XDF signal using a position-dependent effective diffusion coefficient. MIST is derived from the finite-difference form of the Fokker--Planck equation \cite{risken1989fokkerplanck} for paraxial X-ray imaging \cite{paganin2019x,morgan2019applying}; this is a continuity equation that models changes in coherent and diffusive optical flow in an imaging system. MIST solves the Fokker--Planck equation in the context of SBXI, enabling the retrieval of a sample's transmission, phase shift, and XDF. In other words, MIST addresses the multimodal inverse problem associated with the SBXI Fokker--Planck equation. Although the paraxial-optics Fokker--Planck generalization to the transport-of-intensity equation was only realized in 2019 \cite{paganin2019x,morgan2019applying}, the Fokker--Planck equation, in general, is very widely used in various applied mathematics fields to model the dynamics of systems with random processes. Indeed, there are many thousands of research articles relating to the Fokker--Planck equation. Some examples of applications are in hydrodynamics \cite{shea1997fokker,chavanis2003generalized}, gaseous micro-flows \cite{gorji2012kinetic,zhang2019particle}, hot plasmas \cite{cooper1971compton,kolobov2003fokker}, and quantum optics \cite{drummond1980generalised,carmichael2013statistical}. 

The initial formulation of the paraxial-optics Fokker--Planck equation \cite{paganin2019x,morgan2019applying} considered the evolution of an X-ray wavefield as it propagates from the X-ray source through the sample toward the detector plane, a perspective that was adopted by all previously developed MIST variants \cite{pavlov2020x, alloo2021speckle, pavlov2021directional, alloo2022dark, alloo2023m}. Under this perspective, reference speckles will be reduced in intensity, transversely shifted, and blurred out in the presence of a sample due to X-ray attenuation, refraction, and diffuse scattering, respectively. By following the optical flow of these coherent and diffusive processes, the \textit{evolving} SBXI Fokker--Planck equation can be formulated. Beltran \textit{et al.} \cite{beltran2023} recently took a different approach in formulating the forward SBXI Fokker--Planck equation by examining the optical flows in an SBXI system in reverse. They investigated how the sample-plus-speckle image at the detector plane transforms as it propagates from the detector back through the sample, ultimately generating the reference-speckle image. The motivation for this re-formulation was to develop a computationally efficient single-exposure multimodal retrieval algorithm for 2D reference-pattern imaging techniques. The successful experimental demonstration suggests that, indeed, the perspective of reverse-flow can be used to formulate the \textit{devolving} Fokker--Planck equation for SBXI, and this can then be used to solve the associated multimodal inverse problem.

Beltran \textit{et al.} \cite{beltran2023} suggested that although the \textit{evolving} and \textit{devolving} perspectives of the Fokker--Planck equation for SBXI are physically equivalent, they may be inequivalent in terms of the associated inverse problem. The approach and results of Beltran \textit{et al.} initiated the following question: `How would the MIST-recovered multimodal images differ if the algorithms were derived beginning from the devolving paraxial-optics Fokker--Planck equation?'. In this paper, the two perspectives of the Fokker--Planck equation for SBXI are explored alongside the associated inverse problems and recovered multimodal images. We present single- and multiple-exposure multimodal retrieval algorithms using both the evolving and devolving forms of the SBXI Fokker--Planck equation. The various approaches are applied to experimental SBXI data of two samples collected using synchrotron light. It is seen that the devolving perspective allows for the separation of the two main channels of Fokker--Planck XDF contrast, namely, microstructure- and edge-induced XDF. The paper ends with a discussion on possible reasons why differences in the XDF images are seen using the two perspectives and future research avenues are discussed. 
\begin{figure}[!tb]
    \centering
    \includegraphics[width=0.75\textwidth]{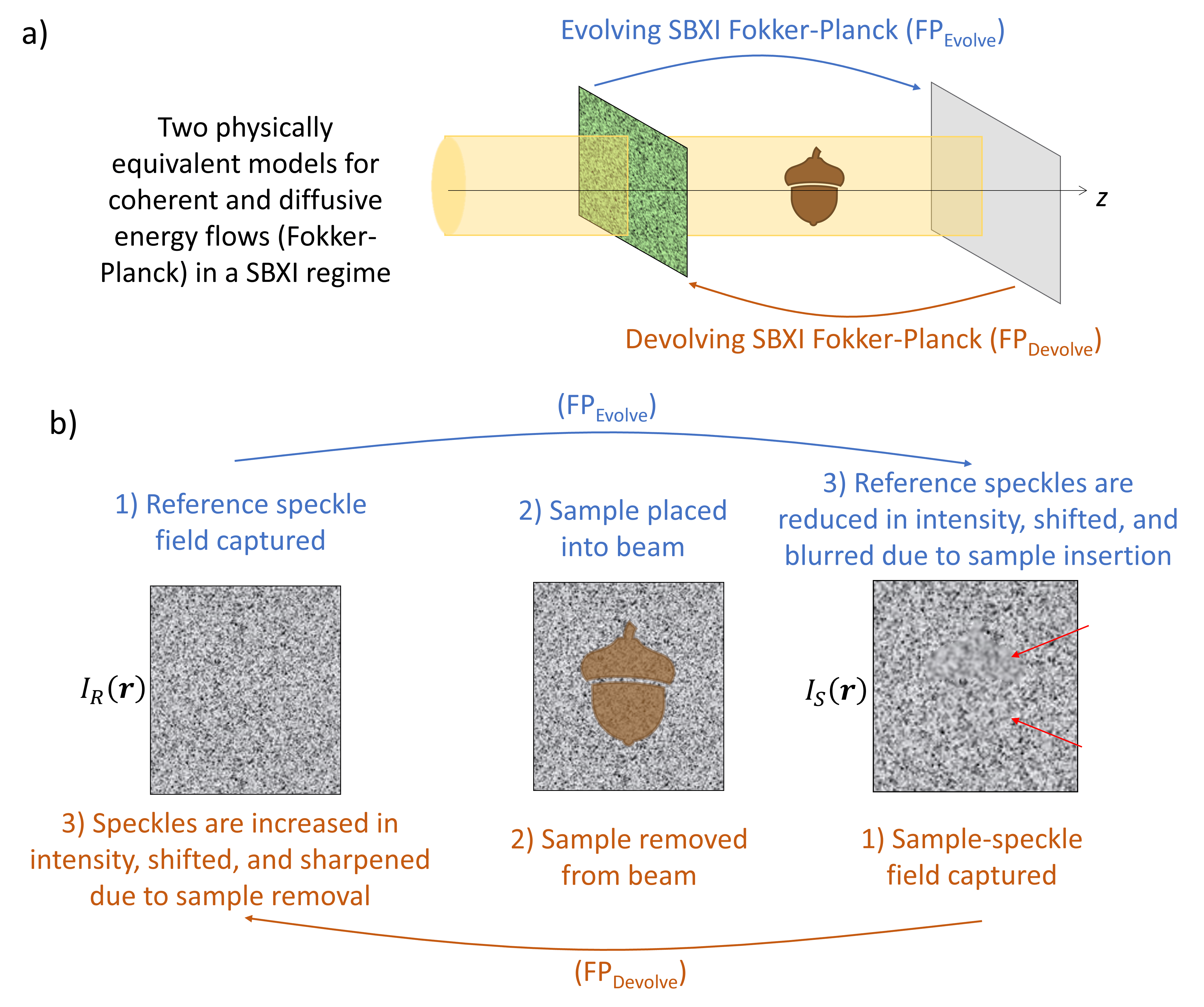}
    \caption{a) Diagram showing the two perspectives of the SBXI Fokker--Planck equation: evolving, $FP_\textrm{Evolve}$, and devolving, $FP_\textrm{Devolve}$. b) SBXI system flowchart illustrating coherent and diffusive speckle flows under each perspective. Red arrows in $I_\textrm{S}(\textbf{r})$ indicate regions where speckles are modified compared to $I_\textrm{R}(\textbf{r})$.}
    \label{fig:DifferentPerspectives}
\end{figure}
\section{Fokker--Planck equation for speckle-based X-ray imaging}
The Fokker--Planck equation is a partial differential equation that models the evolution in position $x$ and time $t$ of a random variable's probability-density distribution, $p(x,t)$, incorporating terms for both drift and diffusion. The one-dimensional (1D) Fokker--Planck equation is a forward Kolmogorov equation and is given by \cite{risken1989fokkerplanck}
\begin{equation}
\label{eqn:FokkerPlanck_General}
\frac{\partial}{\partial t}p(x,t) = - \frac{\partial}{\partial x}\left[\mu(x,t)p(x,t)\right] + \frac{\partial^2}{\partial x^2}\left[\sigma^2(x,t)p(x,t)\right], 
\end{equation}
where $\mu(x,t)$ and $\sigma^2(x,t)$ represent the drift and diffusion coefficients, respectively. Classically, $\sigma^2(x,t)$ in Eq.~\ref{eqn:FokkerPlanck_General} is defined to be strictly positive; $\sigma^2(x,t)>0$ \cite{risken1989fokkerplanck}. When the Fokker--Planck equation is employed in the context of paraxial X-ray imaging, it is considered in 2D corresponding to transverse spatial coordinates $\textbf{r} = (x,y)$, such that the probability distribution $p(\textbf{r},t)$ represents the likelihood of X-ray beamlets hitting a position $\textbf{r}$ on the detector (which corresponds to the X-ray wavefield's intensity), $\mu(\textbf{r},t)$ describes the spatial drift of the beamlets due to sample-induced refraction, and $\sigma^2(\textbf{r},t)$ characterizes the diffusion of X-ray beamlets caused by sample features. The finite-difference form of the Fokker--Planck equation for paraxial X-ray imaging is \cite{paganin2019x,morgan2019applying, paganin2023paraxial} 
\begin{equation}
\label{eqn:FokkerPlanck_Xray}
I(\textbf{r},z=\Delta) = I(\textbf{r},z=0) - \frac{\Delta}{k}\nabla_\perp \cdot \left[I(\textbf{r},z)\nabla_\perp \phi(\textbf{r},z)\right]_{z=0} +\\ \Delta^2\nabla_\perp^2\left[D(\textbf{r}, z=0) I(\textbf{r},z) \right]_{z=0}.
\end{equation}
In the above equation, $I$ is the intensity of the monochromatic paraxial X-ray wavefield, $\phi$ is the phase of the wavefield, $k$ is its wavenumber, $\nabla_\perp$ is the gradient operator in the $\textbf{r} = (x,y)$ planes (i.e., perpendicular to the optical axis $z$), $\nabla_\perp^2$ is the transverse Laplacian operator defined as $\nabla_\perp^2 = (\partial^2/\partial x^2+ \partial^2/\partial y^2)$, and $D = (F \theta^2)/2$ is the dimensionless effective diffusion coefficient which considers the characteristic sample-imposed X-ray diffusion as well as the fraction $F$ of incident X-rays that are diffusely scattered through an angle $\theta$ \cite{paganin2019tutorials,paganin2023paraxial}\footnote{Note that the dimensionless effective diffusion coefficient, denoted by $D(\textbf{r}, z=0)$ in this manuscript, corresponds to $D_\textrm{eff}(\textbf{r}, z=0)$ as used in Ref.~\citenum{paganin2019x} and our prior MIST works \cite{pavlov2020x,alloo2021speckle,alloo2022dark,alloo2023m}. For readability, the subscript `eff' has been omitted in the present work.}. The exit surface of the attenuating, refracting, and diffusing sample is located at the $z=0$ plane, and the intensity evolution of this sample-perturbed wavefield follows Eq.~\ref{eqn:FokkerPlanck_Xray} to the plane $z=\Delta$. Equation~\ref{eqn:FokkerPlanck_Xray} is an energy-conserving equation which states that sample-imposed changes to the $z=0$ intensity distribution at the position of the detector must be due to either coherent or diffuse optical-energy flow. The coherent-flow term separates out the X-ray refraction into a lensing and prism term, which describes the local focussing or defocussing of X-ray intensity (related to $\nabla_\perp^2 \phi$) and the transverse shift in intensity (related to $\nabla_\perp \phi$) \cite{paganin2019x}, respectively. The diffusive-flow term represents the position-dependent local blurring of the X-ray intensity, which is often associated with SAXS due to spatially random unresolved sample microstructure.

Paganin and Morgan initially established Eq.~\ref{eqn:FokkerPlanck_Xray} by considering the evolution of an X-ray wavefield as it propagates from the sample's exit surface \cite{paganin2019x,morgan2019applying}. This \textit{evolving }paraxial-optics Fokker--Planck model was applied in the context of SBXI (see Eq.~55 in Ref.~\citenum{paganin2019x}) and this equation provided the basis for the MIST algorithm \cite{pavlov2020x}. All of the MIST algorithms to date \cite{pavlov2020x, alloo2021speckle, pavlov2021directional, alloo2022dark, alloo2023m} employ the SBXI Fokker--Planck perspective presented in Paganin and Morgan \cite{paganin2019x}. The SBXI Fokker--Planck equation for an attenuating object under this \textit{evolving} perspective is \cite{alloo2021speckle,alloo2022dark}
\begin{equation}
\label{eqn:FPphase_Evolve}
I_\textrm{S}(\textbf{r}) = I_\textrm{R}(\textbf{r})t_{0_\textrm{Ev.}}(\textbf{r}) - \frac{\Delta}{k}\nabla_{\perp}\cdot \left[I_\textrm{R}(\textbf{r})t_{0_\textrm{Ev.}}(\textbf{r})\nabla_{\perp}\phi_{\textrm{Ev.}}(\textbf{r})\right]+\Delta^2\nabla_{\perp}^{2}\left[D_{\textrm{Ev.}}(\textbf{r})I_\textrm{R}(\textbf{r})t_{0_\textrm{Ev.}}(\textbf{r})\right].
\end{equation}
$I_\textrm{R}(\textbf{r})$ and $I_\textrm{S}(\textbf{r})$ denote the reference-speckle and sample-plus-speckle images, respectively, generated by capturing an image of the membrane-modulated X-ray wavefield in the absence and presence of the sample, respectively. $t_0$ denotes the sample's transmission, which is unity where there is no X-ray attenuation and zero where the X-ray beam is fully attenuated. In the case of the pure-phase sample, Eq.~\ref{eqn:FPphase_Evolve} can be simplified by setting $t_0(\textbf{r}) = 1$ \cite{pavlov2020x}. Equation~\ref{eqn:FPphase_Evolve} models how speckles in $I_\textrm{R}(\textbf{r})$ are reduced in intensity, transversely shifted, and blurred out due to placing a sample into the beam, generating $I_\textrm{S}(\textbf{r})$ which is captured a distance $\Delta$ downstream of the sample. This evolving SBXI Fokker--Planck model is demonstrated by the blue annotations in Fig.~\ref{fig:DifferentPerspectives}. The reduction in speckle intensity, transverse speckle shift, and speckle blurring are associated with the sample's transmission $t_{0_\textrm{Ev.}}(\textbf{r})$, phase shift $\phi_{\textrm{Ev.}}(\textbf{r})$, and dimensionless effective diffusion coefficient $D_{\textrm{Ev.}}(\textbf{r})$, respectively. It is this dimensionless effective diffusion coefficient $D_{\textrm{Ev.}}(\textbf{r})$ that quantifies the XDF signal throughout this paper. 

In Eq.~\ref{eqn:FPphase_Evolve} above and the following devolving Fokker--Planck equation (Eq.~\ref{eqn:FPphase_Devolve}), the subscripts `Ev.' and `Dev.' denote the evolving and devolving perspectives, respectively. These subscripts are applied to the transmission, phase, and effective diffusion coefficient to account for the possibility that the solutions to the inverse problem using the two models may be different \cite{beltran2023}, and hence, the recovered multimodal images may also be different.    

Following Beltran \textit{et al.} \cite{beltran2023}, the forward problem of SBXI may also be derived by considering the `reverse optical flow'; this `proceeds from effect to cause' by transforming from $I_\textrm{S}(\textbf{r})$ back through the sample and then towards the speckle-generating membrane to give $I_\textrm{R}(\textbf{r})$. The physically equivalent \textit{devolving} SBXI Fokker--Planck equation for an attenuating object is given in their paper as
\begin{equation}
\label{eqn:FPphase_Devolve}
I_\textrm{R}(\textbf{r}) = \frac{1}{t_{0_\textrm{Dev.}}(\textbf{r})}\left(I_\textrm{S}(\textbf{r}) + \frac{\Delta}{k}\nabla_{\perp}\cdot \left[I_\textrm{S}(\textbf{r})\nabla_{\perp}\phi_{\textrm{Dev.}}(\textbf{r})\right]-\Delta^2\nabla_{\perp}^{2}\left[D_{\textrm{Dev.}}(\textbf{r})I_\textrm{S}(\textbf{r})\right]\right).
\end{equation}
Under this perspective, speckles in $I_\textrm{S}(\textbf{r})$ will increase in intensity, transversely shift, and be locally sharpened due to sample removal from the SBXI system. The orange annotations in Fig.~\ref{fig:DifferentPerspectives} illustrate the devolving SBXI Fokker--Planck model and the relevant speckle mechanisms. It is important to recognize here that Beltran \textit{et al.} wrote down Eq.~\ref{eqn:FPphase_Devolve} from an intuitive viewpoint of the optical flows in SBXI and with analogy to the already-derived SBXI Fokker--Planck equation presented in previous work \cite{paganin2019x,morgan2019applying}.

To derive the form in Eq.~\ref{eqn:FPphase_Devolve}, note that $I_\textrm{S}(\textbf{r})$ may be devolved into $I_\textrm{R}(\textbf{r})$ by first applying the reverse refraction operator 
\begin{equation}
    \hat{\mathcal{D}}_1(\textbf{r}) = 1 + \frac{\Delta}{k}\nabla_{\perp}\cdot \left[\nabla_{\perp}\phi(\textbf{r})\left[\right]\right],
\end{equation}
followed by the inverse-diffusion operator, 
\begin{equation}
    \hat{\mathcal{D}}_2(\textbf{r}) = 1 - \Delta^2\nabla_{\perp}^2 \left[D(\textbf{r})\left[\right]\right],
\end{equation}
and then the inverse-sample-attenuation operator 
\begin{equation}
    \hat{\mathcal{D}}_3(\textbf{r}) = \frac{1}{t_{0}(\textbf{r})}\left[ \right];
\end{equation}
in the preceding three equations, a hat indicates that the symbol represents an operator, and empty square brackets indicate the `slot' into which the operated-upon function is to be placed. With the usual convention that operators act from right to left, so that $\hat{\mathcal{D}}_3\hat{\mathcal{D}}_2\hat{\mathcal{D}}_1f(\textbf{r})$ means $\hat{\mathcal{D}}_3(\hat{\mathcal{D}}_2(\hat{\mathcal{D}}_1(f(\textbf{r}))))$, we may then write the reverse-flow equation as
\begin{equation}
    I_\textrm{R}(\textbf{r}) \approx \hat{\mathcal{D}}_3(\textbf{r})\hat{\mathcal{D}}_2(\textbf{r})\hat{\mathcal{D}}_1(\textbf{r})I_\textrm{S}(\textbf{r}).
\end{equation}
Upon expanding out the operator products and retaining only those terms up to the second order in $\Delta$, Eq.~\ref{eqn:FPphase_Devolve} is obtained. 

\section{The multimodal inverse problem}\label{The multimodal inverse problem}
The SBXI multimodal inverse problem, in the context of Fokker--Planck-based algorithms, involves recovering an object's transmission $t_0(\textbf{r})$, phase shift $\phi$(\textbf{r}), and dimensionless effective diffusion coefficient $D(\textbf{r}; \Delta)$ given some algorithm-appropriate SBXI data $I_\textrm{R}(\textbf{r})$ and $I_\textrm{S}(\textbf{r})$. The term `algorithm-appropriate' highlights that some algorithms require a single set of $I_\textrm{R}(\textbf{r})$ and $I_\textrm{S}(\textbf{r})$, while some require multiple sets that are acquired by moving the speckle membrane to new transverse positions and recording $I_\textrm{R}(\textbf{r})$ and $I_\textrm{S}(\textbf{r})$ pairs at each step. To investigate the key differences in the associated inverse problems of the \textit{evolving} and \textit{devolving} SBXI Fokker--Planck equations, we compare the perspectives using single- and multiple-exposure approaches. The mathematical procedure outlined in Beltran \textit{et al.} \cite{beltran2023} for multimodal signal extraction using a single set of SBXI data will be investigated. These authors used the devolving SBXI Fokker--Planck equation (Eq.~\ref{eqn:FPphase_Devolve}), and here we present their devolving perspective-based algorithm and compare with one that we derive that instead begins from the evolving equation (Eq.~\ref{eqn:FPphase_Evolve}). The multiple-exposure MIST approach in Alloo \textit{et al.} \cite{alloo2023m}, which begins from the evolving SBXI Fokker--Planck equation, will be compared with that derived using similar mathematical methods but beginning from the devolving Fokker--Planck equation.

\subsection{Single-exposure approach from the evolving perspective of SBXI}\label{Single_Evolve}
To derive a single-exposure SBXI multimodal retrieval algorithm from the evolving Fokker--Planck equation, we follow Beltran \textit{et al.}'s \cite{beltran2023} theoretical approach, which considered the devolving equation. Beginning from Eq.~\ref{eqn:FPphase_Evolve}, we assume that the sample is homogeneous such that $\gamma = \delta/\beta$ is approximately constant throughout the sample's entire volume, where $\delta$ and $\beta$ are the real and imaginary components of a material's X-ray refractive index $1-\delta(\textbf{r})+i\beta(\textbf{r})$, describing X-ray refraction and attenuation, respectively. First, the sample's transmission term $t_0(\textbf{r})$ is calculated using the transport-of-intensity equation-based approach for SBXI developed in Pavlov \textit{et al.} \cite{pavlov2020single} (cf. Eq.~18 therein). A sample's transmission image can be retrieved using
\begin{equation}
\label{eqn:Pavlov_t0}
t_0 \approx \mathcal{F}^{-1}\left[\frac{1}{1+\frac{\Delta\gamma}{2k}\left(k_x^2 +k_y^2\right)} \mathcal{F}\left[ \frac{I_\textrm{S}}{I_\textrm{R}}\right]\right].
\end{equation}
Above, $\mathcal{F}$ denotes the two-dimensional Fourier transform with respect to $x$ and $y$, $k_x$ and $k_y$ are the corresponding Fourier-space variables, and explicit functional dependence on \textbf{r} is here and henceforth dropped (for clarity). Pavlov \textit{et al.}'s transport-of-intensity-based approach neglects sample-imposed X-ray diffusion, however, it provides a good first approximation to $t_0$ for weakly to moderately diffusing samples; if necessary, $t_0$ could always be iteratively refined afterwards, as suggested in Ref.~\citenum{beltran2023}.

The phase shift in Eq.~\ref{eqn:FPphase_Evolve} can be written in terms of the recovered transmission image using 
\begin{equation}
\label{eqn:Phase_and_t0}
\phi = \frac{\gamma}{2} \textrm{ln}(t_0).
\end{equation}
The coherent-flow term in Eq.~\ref{eqn:FPphase_Evolve} can then be evaluated by using the relationship in Eq.~\ref{eqn:Phase_and_t0} and expanding out the divergence operator via the identity $\nabla_{\perp}\cdot \left[A\nabla_{\perp}B\right] = \left(\partial_x\left[A\partial_x B\right] + \partial_y\left[A\partial_y B\right]\right)$ \cite{spiegel1959vector} for two scalar functions $A$ and $B$, where $\partial_x$ and $\partial_y$ denote the partial derivative operators with respect to $x$ and $y$, respectively.
Using these techniques, the coherent-flow term in Eq.~\ref{eqn:FPphase_Evolve} can be written as
\begin{equation}
\label{eqn:FPphase_Evolve_CoherentExpand}
\frac{\Delta}{k}\nabla_{\perp}\cdot \left[I_\textrm{R}t_{0_\textrm{Ev.}}\nabla_{\perp}\phi_{\textrm{Ev.}}\right] = \frac{\Delta\gamma}{2k}\left(\partial_x\left[I_\textrm{R}t_{0_\textrm{Ev.}}\partial_x \textrm{ln}(t_0)\right] + \partial_y\left[I_\textrm{R}t_{0_\textrm{Ev.}}\partial_y \textrm{ln}(t_0)\right]\right).
\end{equation}
The derivative operators above can be evaluated using the Fourier derivative theorem \cite{paganin2006coherent}: $\partial_{x, y} =  \mathcal{F}^{-1}ik_{x,y}\mathcal{F}$. Using this technique, the coherent-flow term can be evaluated using the equation above, the XDF (as quantified by the dimensionless effective diffusion coefficient) can be recovered by rearranging Eq.~\ref{eqn:FPphase_Evolve} for $D_{\textrm{Ev.}}$, which gives
\begin{equation}
\label{eqn:FPphase_Evolve_DF}
D_{\textrm{Ev.}} = \frac{\nabla_{\perp}^{-2}\left[I_\textrm{S} + \frac{\Delta\gamma}{2k}\left(\partial_x\left[I_\textrm{R}t_{0_\textrm{Ev.}}\partial_x \textrm{ln}(t_0)\right] + \partial_y\left[I_\textrm{R}t_{0_\textrm{Ev.}}\partial_y \textrm{ln}(t_0)\right]\right)-I_\textrm{R}t_{0_\textrm{Ev.}}\right]}{\Delta^2I_\textrm{R}t_{0_\textrm{Ev.}}}.
\end{equation}
The inverse transverse Laplacian operator $\nabla_{\perp}^{-2}$ can be evaluated using \cite{paganin2006coherent}
\begin{equation}
\label{eqn:InverseLaplacian}
\nabla_{\perp}^{-2} = -\mathcal{F}\frac{1}{(k_x^2+k_y^2)+\epsilon}\mathcal{F},
\end{equation}
where $\epsilon$ is a small positive regularization parameter \cite{tikhonov1977solutions} used to stabilize the numerical instability near the Fourier-space origin. 

\subsection{Single-exposure approach from the devolving perspective of SBXI}\label{Single_Devolve}
The single-exposure devolving approach described in this section is that of Beltran \textit{et al.} \cite{beltran2023}. Similar to the evolving variant, the transmission image, and hence phase image, is first calculated using the method of Pavlov \textit{et al.} \cite{pavlov2020single} -- see Eq.~\ref{eqn:Pavlov_t0} above. Thus, the recovered transmission and phase images will be the same for the two single-exposure approaches investigated in this work, namely, $t_{0_\textrm{Ev.}} = t_{0_\textrm{Dev.}} \equiv t_{0}$.

To recover the devolving-XDF signal, the devolving form of the SBXI Fokker--Planck equation (Eq.~\ref{eqn:FPphase_Devolve}) is rearranged for $D_{\textrm{Dev.}}$, hence
\begin{equation}
\label{eqn:FPphase_Devolve_DF}
D_{\textrm{Dev.}} = \frac{\nabla_{\perp}^{-2}\left[I_\textrm{S} + \frac{\Delta\gamma}{2k}\left(\partial_x\left[I_\textrm{S}\partial_x \textrm{ln}(t_0)\right] + \partial_y\left[I_\textrm{S}\partial_y \textrm{ln}(t_0)\right]\right)-I_\textrm{R}t_{0_\textrm{Dev.}}\right]}{\Delta^2I_\textrm{S}},
\end{equation}
where the inverse transverse Laplacian operator can be evaluated using Eq.~\ref{eqn:InverseLaplacian}. 

\subsection{Multiple-exposure approach from the evolving perspective of SBXI}\label{Multi_Evolve}
The two single-exposure approaches above recover the sample's transmission and phase signals in an identical manner, as given by Eq.~\ref{eqn:Pavlov_t0}. Thus, to investigate how multimodal images (transmission, phase, and XDF) are influenced by the perspective of the forward problem, we also investigate multiple-exposure approaches. The additional speckle positions required in these multiple-exposure algorithms also improve the spatial resolution of the retrieved images, as seen in the results section. We take the MIST approach in Alloo \textit{et al.} \cite{alloo2023m} and apply the mathematical procedures to the evolving and devolving Fokker--Planck equation.

We first present the approach developed in Ref.~\citenum{alloo2023m}, which used the evolving SBXI Fokker--Planck as its forward model. The approach has the same assumption as the single-exposure approaches above, namely that the sample has a constant $\gamma = \delta/\beta$ throughout its volume. Furthermore, it begins with the phase-object form of Eq.~\ref{eqn:FPphase_Evolve} (setting $t_{0_\textrm{Ev.}} =1$). The coherent-flow term is simplified by neglecting the prism term \cite{paganin2019x}, so that $\nabla_{\perp}I_\textrm{R} \cdot \nabla_{\perp}\phi_\textrm{Ev.} \approx 0$. This assumption comes from the fact that $\phi$ is spatially slowly varying (this assumption was employed in Pavlov \textit{et al.} \cite{pavlov2020single}). Using this assumption and expanding out the diffusive optical flow term using the identity $\nabla_\perp^2[AB] = A\nabla_\perp^2B + B\nabla_\perp^2A +2\nabla_\perp A\cdot\nabla_\perp B$ \cite{spiegel1959vector} for two scalar functions $A$ and $B$ gives Eq.~\ref{eqn:FPphase_Evolve} in the form
\begin{equation}
\label{eqn:FPphase_Evolve_Alt}
I_\textrm{S} = I_\textrm{R} + \Delta I_\textrm{R}\nabla_{\perp}^2\left[\frac{-1}{k}\phi_{\textrm{Ev.}} + \Delta D_{\textrm{Ev.}}^\textrm{Phase}\right]+\Delta^2D_{\textrm{Ev.}}^\textrm{Phase}\nabla_{\perp}^2I_\textrm{R} +\\2\Delta^2\partial_xI_\textrm{R}\partial_xD_{\textrm{Ev.}}^\textrm{Phase}+2\Delta^2\partial_yI_\textrm{R}\partial_yD_{\textrm{Ev.}}^\textrm{Phase}.
\end{equation}
The superscript \textit{Phase} on all effective diffusion coefficient terms above indicates that a phase object is assumed in the forward model for this algorithm. This superscript is later omitted as the approach is extended to weakly attenuating objects. Equation~\ref{eqn:FPphase_Evolve_Alt} is a linear equation in terms of four unknowns $[(-1/k)\phi_{\textrm{Ev.}} + \Delta D_{\textrm{Ev.}}^\textrm{Phase}]$, $D_{\textrm{Ev.}}^\textrm{Phase}$, $\partial_xD_{\textrm{Ev.}}^\textrm{Phase}$, and $\partial_yD_{\textrm{Ev.}}^\textrm{Phase}$. Four, or more, instances of Eq.~\ref{eqn:FPphase_Evolve_Alt} can be generated by utilizing several sets of SBXI data (an $I_\textrm{R}$ and $I_\textrm{S}$ pair). The resultant system of linear equations (over-determined system if more than four SBXI speckle positions are used) can then be solved using a Tikhonov-regularized \cite{tikhonov1977solutions} QR decomposition with a suitable regularization parameter, as explained in Alloo \textit{et al.} \cite{alloo2023m}. A reasonable value for the optimal Tikhonov regularization parameter for solving this system of equations is given by the standard deviation of the coefficient matrix divided by 10$^4$, as determined in Ref.~\citenum{alloo2023m}, where this value optimized the computed solutions' image quality. To reconstruct the object's true effective diffusion coefficient, the $D_{\textrm{Ev.}}^\textrm{Phase}$, $\partial_xD_{\textrm{Ev.}}^\textrm{Phase}$, and $\partial_yD_{\textrm{Ev.}}^\textrm{Phase}$ solutions from the linear system can be suitably aggregated using
\begin{equation}
\label{eqn:FPphase_Evolve_DFAgg}
\tilde{D}_\textrm{Ev.}^\textrm{Phase} = \mathcal{F}^{-1}\left[\textrm{e}^{-\rho\left(k_x^2 + k_y^2\right)}\mathcal{F}\left(D_{\textrm{Ev.}}^\textrm{Phase}\right)+ \frac{1-\textrm{e}^{-\rho\left(k_x^2 + k_y^2\right)}}{ik_x - k_y}\mathcal{F}\left(\partial_xD_{\textrm{Ev.}}^\textrm{Phase} + i\partial_yD_{\textrm{Ev.}}^\textrm{Phase} \right)\right],
\end{equation}
where $\rho$ is the cut-off parameter which defines how much of each solution contributes to the final true dimensionless effective diffusion coefficient image. A tilde has been placed over $D_\textrm{Ev.}^\textrm{Phase}$ in the above equation to indicate that it represents the solution-aggregated effective diffusion coefficient, providing an accurate reconstruction of the object's XDF. The optimal value of $\rho$ can be chosen by optimizing the image quality of $\tilde{D}_\textrm{Ev.}^\textrm{Phase}$ for different values of $\rho$, for example, by assessing the signal-to-noise ratio (SNR) or naturalness image quality evaluator (NIQE) \cite{mittal2012making}.

The object's phase shift is retrieved by using the recovered $\tilde{D}_\textrm{Ev.}^\textrm{Phase}$ in the evolving Fokker--Planck equation which neglects the prism term $\nabla_{\perp}I_\textrm{R} \cdot \nabla_{\perp}\phi_\textrm{Ev.}$, and then rearranging to give
\begin{equation}
\label{eqn:FPphase_Evolve_Phi}
\phi_{\textrm{Ev.}} = \nabla_\perp^{-2}\left[\frac{k\left(I_\textrm{R}-I_\textrm{S}+\Delta^2\nabla_\perp^{2}\left[\tilde{D}_\textrm{Ev.}^\textrm{Phase}I_\textrm{R} \right]\right)}{\Delta I_\textrm{R}}\right].
\end{equation}
Above, the inverse transverse Laplacian operator can be applied by using Eq.~\ref{eqn:InverseLaplacian}. Recall that a non-attenuating object has been assumed thus far in the derivation. To extend to a weakly attenuating object, the transmission term can be calculated by rearranging Eq.~\ref{eqn:Phase_and_t0} and using the recovered $\phi_{\textrm{Ev.}}$ within, after which the attenuation-corrected dimensionless effective diffusion coefficient can be calculated simply by dividing the result of Eq.~\ref{eqn:FPphase_Evolve_DFAgg} by $t_{0_\textrm{Ev.}}$. This result was derived in Alloo \textit{et al.} \cite{alloo2021speckle, alloo2022dark}. Explicitly, 
\begin{equation}
\label{eqn:FPphase_Evolve_DFAtten}
\tilde{D}_\textrm{Ev.} = \frac{\tilde{D}_\textrm{Ev.}^\textrm{Phase}}{t_{0_\textrm{Ev.}}}=\tilde{D}_\textrm{Ev.}^\textrm{Phase}\,\textrm{e}^{\frac{-2}{\gamma}\phi_\textrm{Ev.}}.
\end{equation}

\subsection{Multiple-exposure approach from the devolving perspective of SBXI}\label{Multi_Devolve}
To achieve a devolving multiple-exposure algorithm, we begin by following the approach above, with the phase-object version of the devolving SBXI Fokker--Planck equation (i.e., set $t_{0_\textrm{Dev.}} = 1$ in Eq.~\ref{eqn:FPphase_Devolve}). The prism term in Eq.~\ref{eqn:FPphase_Devolve} is $\nabla_{\perp}I_\textrm{S} \cdot\nabla_{\perp}\phi_\textrm{Dev.}$, and as in the previous section, this is neglected. Employing this approximation and expanding the diffusion term of the devolving equation similarly to Eq.~\ref{eqn:FPphase_Evolve_Alt}, we arrive at
\begin{equation}
\label{eqn:FPphase_Devolve_ALT}
I_\textrm{R} = I_\textrm{S} + \Delta I_\textrm{S}\nabla_{\perp}^2\left[\frac{1}{k}\phi_{\textrm{Dev.}} - \Delta D_{\textrm{Dev.}}^\textrm{Phase} \right]-
\Delta^2D_{\textrm{Dev.}}^\textrm{Phase}\nabla_{\perp}^2I_\textrm{S} -\\2\Delta^2\partial_xI_\textrm{S}\partial_xD_{\textrm{Dev.}}^\textrm{Phase}-2\Delta^2\partial_yI_\textrm{S}\partial_yD_{\textrm{Dev.}}^\textrm{Phase}.
\end{equation}
Equation~\ref{eqn:FPphase_Devolve_ALT} is a linear equation in terms of $[(1/k)\phi_{\textrm{Dev.}} - \Delta D_{\textrm{Dev.}}^\textrm{Phase}]$, $D_{\textrm{Dev.}}^\textrm{Phase}$, $\partial_xD_{\textrm{Dev.}}^\textrm{Phase}$, and $\partial_yD_{\textrm{Dev.}}^\textrm{Phase}$, hence the generated system for different $I_\textrm{R}$ and $I_\textrm{S}$ pairs can be solved using the Tikhonov-regularized QR decomposition described in the previous section. The corresponding devolving dimensionless effective diffusion coefficient, $\tilde{D}_\textrm{Dev.}^\textrm{Phase}$, can be calculated using an equivalent form of Eq.~\ref{eqn:FPphase_Evolve_DFAgg}, but with every occurrence of $D_{\textrm{Ev.}}^\textrm{Phase}$ replaced with $D_{\textrm{Dev.}}^\textrm{Phase}$, which are the devolving effective diffusion coefficient-related terms retrieved from solving the linear system of equations generated by instances of Eq.~\ref{eqn:FPphase_Devolve_ALT}.

The object's phase shift is given by solving Eq.~\ref{eqn:FPphase_Devolve_ALT} for $\phi_{\textrm{Dev.}} $ and utilizing the retrieved $\tilde{D}_\textrm{Dev.}^\textrm{Phase}$, hence
\begin{equation}
\label{eqn:FPphase_Devolve_Phi}
\phi_{\textrm{Dev.}} = \nabla_\perp^{-2}\left[\frac{k\left(I_\textrm{R}-I_\textrm{S}+\Delta^2\nabla_\perp^{2}\left[\tilde{D}_\textrm{Dev.}^\textrm{Phase}I_\textrm{S} \right]\right)}{\Delta I_\textrm{S}}\right].
\end{equation}
A form of Eq.~\ref{eqn:FPphase_Evolve_DFAtten}, but in terms of the devolving-retrieved signals, can be used to recover the weakly attenuating-object approximation for the dimensionless effective diffusion coefficient $\tilde{D}_\textrm{Dev.}$. 


\section{Experimental methods}
To investigate the retrieved images from the evolving and devolving SBXI Fokker--Planck equations, the four preceding solutions to the multimodal inverse problem were applied to SBXI data sets collected from two samples using synchrotron light: (1) a phantom sample, called the `four-rod' sample, composed of a reed diffuser stick, polymethyl methacrylate (PMMA) rod, toothpick, and tree twig (shown in Fig.~\ref{fig:SBXI_Setup}) and (2) a red currant sample.

The four-rod sample was imaged using SBXI at the Micro-CT beamline at the Australian Synchrotron. The experimental set-up was established in the Micro-CT's first experimental hutch, 24.0 m from the bending magnet source point. A 25 keV monochromatic X-ray beam and a 40 ms exposure time were used. The monochromatic beam had a spectral bandwidth of $\Delta E/E \approx 3\times10^{-2}$, generated by using a double-multilayer monochromator. The sample-to-detector propagation distance was $\Delta$ = 0.7 m. The detector system was a pco.edge 5.5 complementary metal-oxide-semiconductor camera with 2560$\times$2160 pixels, each 6.5 $\times$ 6.5 $\SI{}{\micro\meter}$ in size, coupled to a GGG:Eu/Tb scintillator with a 1 $\times$ optical lens placed in between \cite{arhatari2023micro}. The speckle-generating membrane was two layers of P800 grit sandpaper, positioned 0.3 m upstream of the sample. The generated speckle field had an effective speckle size of 18.3 $\SI{}{\micro\meter}$ and a Michelson visibility \cite{michelson1995studies} of 0.28 at the detector plane. The effective speckle size was measured by taking the average full-width at half-maximum of the autocorrelation function \cite{goodman2020speckle} in the horizontal and vertical directions of the reference-speckle image. The speckle membrane was translated perpendicular to the optical axis to collect a total of 13 SBXI data pairs -- 13 reference-speckle images and 13 corresponding sample-plus-speckle images.

The red currant sample was imaged at the BMO5 beamline at the European Synchrotron Radiation Facility (ESRF). These data were initially acquired and published by Berujon and Ziegler \cite{berujon2016x}. An X-ray energy of 17 keV and spectral bandwidth of $\Delta E/E \approx 10^{-4}$ was produced using a double crystal Si(111) monochromator located 27 m from the X-ray source. The red currant was positioned 55 m from the source, and a piece of sandpaper with grit size P800 was placed 0.5 m upstream of the sample. The detector system was $\Delta = 1$ m downstream of the sample and consisted of a Fast Read-Out Low-Noise (FReLoN) e2V camera coupled to an optical imaging scintillator. The effective pixel size of the optical system was 5.8 $\SI{}{\micro\meter}$, and the sandpaper-generated speckle field had an effective speckle size of 20.4 $\SI{}{\micro\meter}$ and a Michelson visibility of 0.50. A 600 ms X-ray exposure was used to capture 7 pairs ($I_\textrm{R}$ and $I_\textrm{S}$) of SBXI data.

\section{Experimental results}
This section will present the XDF and transmission images for both samples that were retrieved using the four algorithms derived in Sec.~\ref{The multimodal inverse problem}. For the four-rod sample, $\gamma$ was taken as that of PMMA at 25 keV, with $\gamma_\textrm{PMMA} = 2355$, since wood types generally have a similar $\gamma$ value to that of PMMA; cf. the $\gamma$ value used for the wood reconstructions in Alloo \textit{et al.} \cite{alloo2022dark}. Furthermore, the wood--air interfaces in the phase-retrieved images shown later in this manuscript are reconstructed sharply, with no over-smoothing or residual phase contrast, indicating that indeed the value of $\gamma_\textrm{PMMA} = 2355$ is suitable for this type of sample. For the red currant sample, the $\gamma$ value was taken as that of water at 17 keV, $\gamma_\textrm{Currant} = 1146$, which is typical for biological and organic samples. For the multiple-exposure approaches, the entirety of the SBXI data collected for each sample was used: thirteen and seven speckle positions for the four-rod and red currant, respectively. The regularization parameter for the Tikhonov-regularized QR decomposition was selected as recommended in Sec.~\ref{The multimodal inverse problem}, while the regularization parameter for the inverse transverse Laplacian operator for all of the phase reconstructions was set to $\epsilon = 0.0001$ for both samples, following the approach in Ref.~\citenum{alloo2023m}. The cut-off parameter required for the multiple-exposure approach (Eq.~\ref{eqn:FPphase_Evolve_DFAgg}) was $\rho = 10$ $\SI{}{\micro\meter}^2$ and $\rho = 21$ $\SI{}{\micro\meter}^2$ for the four-rod sample and red currant, respectively. 

When comparing the solutions to the two Fokker--Planck equations' inverse problems, the single- and multiple-exposure approaches yield similar results, with the multiple-exposure approaches retrieving higher-resolution images. The single- and multiple-exposure algorithms are based on different mathematical assumptions, yet their retrieved signals converge when comparing the evolving and devolving Fokker--Planck perspectives. As both the evolving and devolving single-exposure algorithms utilize the phase retrieval method from Pavlov \textit{et al.} \cite{pavlov2020single}, as suggested by Beltran \textit{et al.} \cite{beltran2023}, there will be no difference in the retrieved transmission signal between the two single-exposure approaches. However, there can be a difference in the retrieved transmission signal between the two multiple exposure approaches. 

The multimodal reconstructions of the four-rod and red currant presented in this work, along with the Python scripts used to compute these images, are available for download from an open GitHub repository \cite{reconstruction_Github}.

\subsection{Four-rod sample}
\begin{figure}[tb!]
    \centering 
    \includegraphics[width=\linewidth]{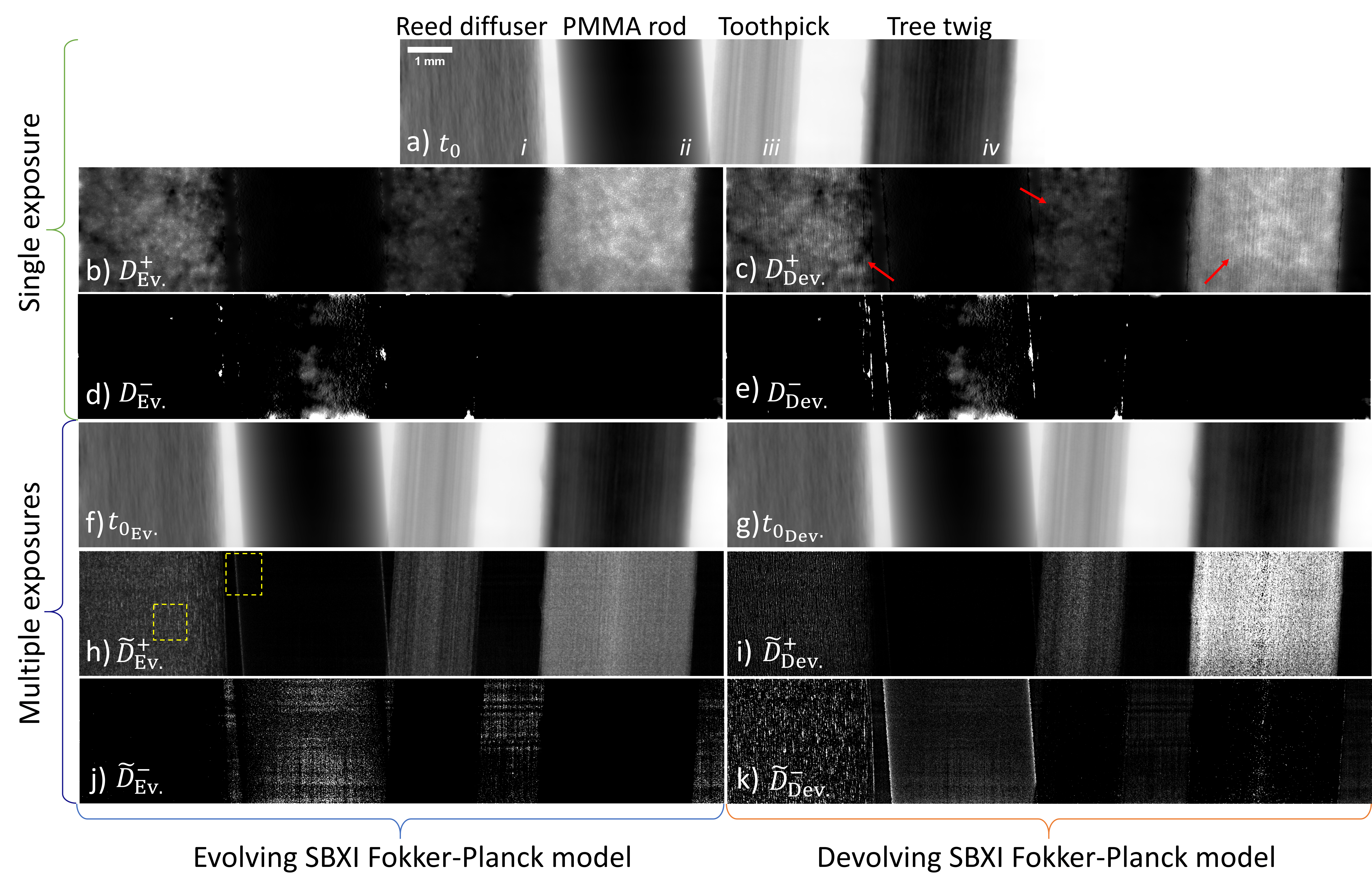} 
    \caption{Recovered transmission and X-ray dark-field (XDF) images of the four-rod sample---\textit{i}) a reed diffuser stick, \textit{ii}) PMMA rod, \textit{iii}) toothpick, and \textit{iv}) a tree twig---using the evolving (left column) and devolving (right column) SBXI Fokker--Planck models. a)--e) were calculated using the single-exposure approaches and f)--k) with the multiple-exposure approaches (13 sets of SBXI data). The red arrows in c) highlight fiber-like structures visible in the devolving-retrieved single-exposure XDF, but not in the evolving-retrieved single-exposure XDF. The yellow-dashed boxes in f) denote the regions of interest for the magnified images shown in Fig.~\ref{fig:FourWood_MultiLineProfile2}. Images are shown in linear grayscale with min and max values of [min, max]: a) = f) = g) [0.8, 1.0], b) = c) = h) = i) [0.0, 8.0]$\times 10^{-11}$, d) = e) = j) [0.0, 1.3]$\times 10^{-12}$, and k) [0.0, 1.3]$\times 10^{-11}$.} 
    \label{fig:FourWood_Total} 
\end{figure}

The four-rod's retrieved XDF and transmission signals using the various approaches are shown in Fig.~\ref{fig:FourWood_Total}. We show explicitly the magnitude of positive and negative components of the reconstructed XDF signals as they appear to contain different information under the two Fokker--Planck perspectives. Positive images were recovered by setting all XDF values below zero to zero, while the negative component was isolated by doing the reverse. The magnitudes of the negative components are shown; thus, a brighter/whiter signal indicates larger negative XDF values. The reconstructed signals that use just one SBXI speckle position, Figs.~\ref{fig:FourWood_Total}a)--e), are expected to have a lower spatial resolution than those reconstructed with multiple sets, Figs.~\ref{fig:FourWood_Total}f)--k). Below, we first identify some interesting features of the recovered single-exposure images, which will then be further demonstrated in the images retrieved when using multiple speckle positions.

The transmission signal in Fig.~\ref{fig:FourWood_Total}a) reveals that the PMMA rod (\textit{ii}) and tree twig (\textit{iv}) attenuate the beam the most in the four-rod sample. Some vertical fibers in the reed diffuser (\textit{i}), toothpick (\textit{iii}), and tree twig (\textit{iv}) can be resolved in the transmission reconstruction. Qualitatively, for low-spatial-frequency features, the reconstructed positive XDF signals are similar using the two evolution models, as seen in Figs.~\ref{fig:FourWood_Total}b) and c); for example, the interior of the reed diffuser stick (\textit{i}) and the tree twig (\textit{iv}) give a strong positive XDF signal. However, noticeable differences are observed for high-spatial-frequency features, such as the rapidly varying just-resolved fibers in the reed diffuser stick (\textit{i}) and the edges of all four materials in the four-rod sample. The negative component of the devolving-retrieved XDF contains information about spatially rapidly varying sample structures, such as sharp edges; some fibers in the reed diffuser (\textit{i}) and the edges of the PMMA rod (\textit{ii}) can be resolved in Fig.~\ref{fig:FourWood_Total}e). In contrast, the $D_{\textrm{Ev.}}^{-}$ image in Fig.~\ref{fig:FourWood_Total}d) appears to contain no useful sample information. An important result seen in the single-exposure results is the recovery of fiber-like structures in the XDF signal, like those indicated by the red arrows in Fig.~\ref{fig:FourWood_Total}c). Notably, when using data from a single speckle position, these structures are only discernible in the devolving-retrieved XDF, Fig.~\ref{fig:FourWood_Total}c), and not in the evolving-retrieved XDF, Fig.~\ref{fig:FourWood_Total}b). The evolving single-exposure XDF contains little information about sample edges, resulting in a smoother XDF. For single-exposure approaches, detailed information about rapidly varying sharp sample edges is only accessible when the devolving Fokker--Planck model is used in the forward problem. The single-exposure results may be summarized as follows: (1) All sample-imposed diffusion of the X-ray illumination manifests as a positive signal under the evolving Fokker--Planck perspective. (2) The negative component of the evolving-retrieved XDF does not clearly convey sample information. (3) Sharp edges are reconstructed with a negative XDF signal using the devolving perspective. (4) The evolving- and devolving-retrieved XDF images converge for low-spatial-frequency structures, such as regions with dense, unresolved microstructure.

The preceding key findings are further reinforced by the four-rod's multiple-exposure reconstructions, which are shown in Figs.~\ref{fig:FourWood_Total}f)--k). The apparent ability of the devolving SBXI Fokker--Planck model to separate out sharp-edge-induced XDF contrast is clearly evident in the $\tilde{D}_{\textrm{Dev.}}^{-}$ image shown in Fig.~\ref{fig:FourWood_Total}k), which was recovered using the devolving multiple-exposure approach. Take the edge of the PMMA rod (\textit{ii}) as one example; there is no XDF contrast (positive or negative) inside the rod, as expected, as it is a homogeneous/smooth material, however, the sharp edges between the rod and air generate a strong negative XDF signal in Fig.~\ref{fig:FourWood_Total}k). This is also seen at the edges of the resolved vertical fibers (i.e., these fibers can be seen in the recovered transmission signal) in the reed diffuser stick (\textit{i}). This also makes the individual fibers more discernible in the devolving-retrieved XDF images compared to the evolving-retrieved ones, where the XDF signal is entirely contained in the positive component of the evolving XDF and hence the fiber-induced signal is overlaid on the reed stick's global background XDF signal. The edge XDF signal observed in our reconstructions has also been reported in the previous MIST algorithms \cite{pavlov2020x,alloo2021speckle, alloo2022dark,pavlov2021directional,alloo2023m}. Notably, these algorithms adopt the evolving Fokker--Planck perspective, and hence consistently show a positive XDF signal at sample edges (cf. any of the XDF reconstructions in Refs.~\citenum{pavlov2020x,alloo2021speckle, alloo2022dark,pavlov2021directional,alloo2023m}). Our multiple-exposure evolving XDF reconstructions of the four-rod sample, Figs.~\ref{fig:FourWood_Total}h) and j), demonstrate that the positive component of the evolving XDF captures both the sharp-edge contribution and the spatially random microstructure diffuse-scatter contribution. In contrast, the devolving Fokker–Planck algorithms separate the sharp-edge XDF contribution into the negative component of the recovered XDF, which is shown clearly in Fig.~\ref{fig:FourWood_Total}k), while the spatially random microstructure diffuse-scatter contribution is retained in the positive component, as shown in Fig.~\ref{fig:FourWood_Total}i).
\begin{figure}[t]
    \centering 
    \includegraphics[width=0.9\linewidth, trim=5 5 5 5,clip]{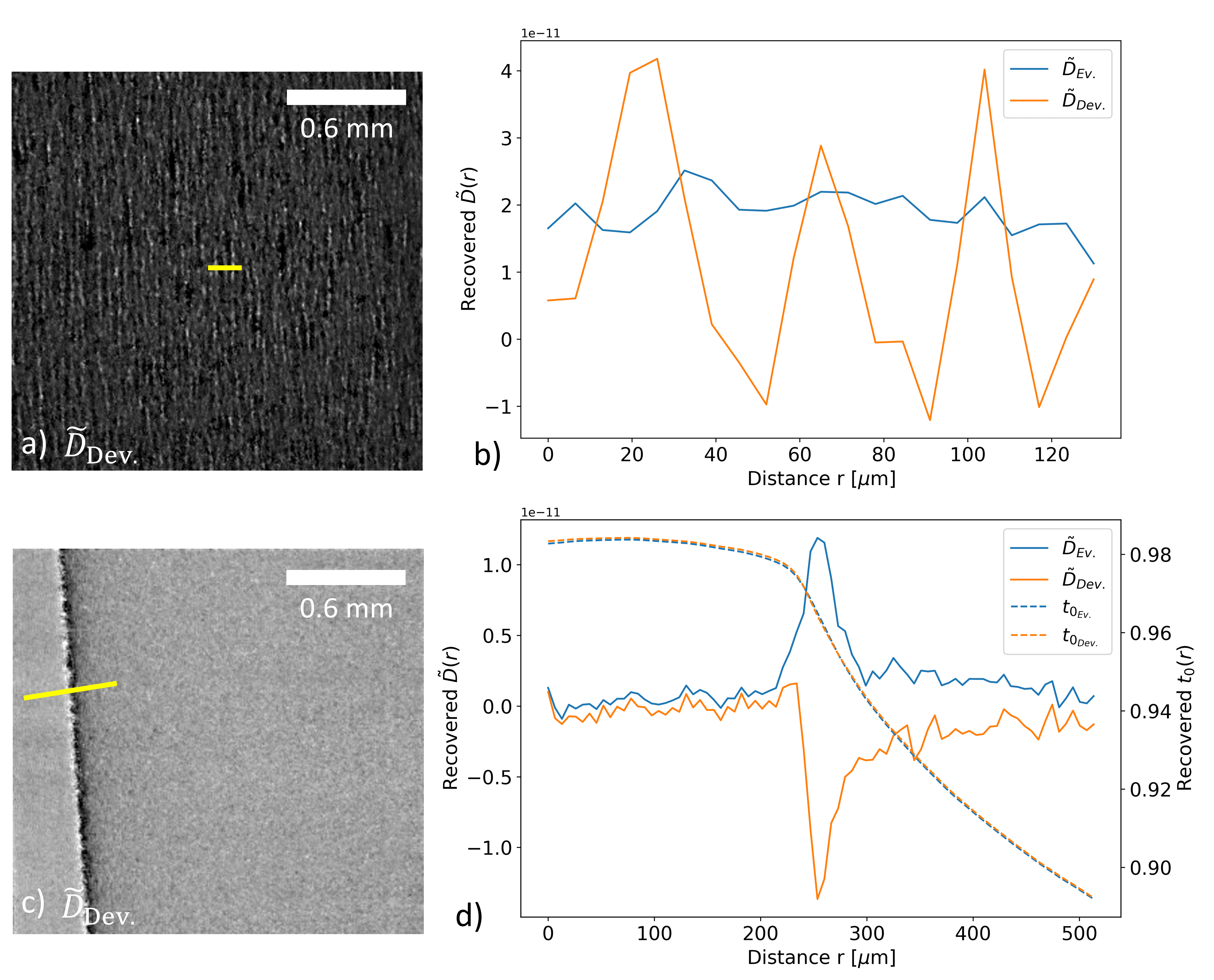} 
    \caption{a) Magnified image of the reed diffuser stick's (\textit{i}) multiple-exposure $\tilde{D}_\textrm{Dev.}$ signal, the exact region is shown by the left-most yellow box in Fig.~\ref{fig:FourWood_Total}h). The yellow line denotes where the line profile shown in b) was traced. b) Retrieved XDF signals across the yellow trace in a). c) Magnified image of the PMMA rod's (\textit{ii}) multiple-exposure $\tilde{D}_\textrm{Dev.}$ signal, the exact region is shown by the right-most yellow box in Fig.~\ref{fig:FourWood_Total}h). The yellow line denotes where the line profile shown in d) was traced. d) Retrieved XDF and transmission signals across the air--PMMA interface, indicated in c). Images are shown in linear grayscale with min and max values of [min, max]: a) [-1.4, 9.7]$\times10^{-11}$ and c) [-0.1, 6.6]$\times10^{-12}$.}
    \label{fig:FourWood_MultiLineProfile2} 
\end{figure}

Magnified images of the multiple-exposure devolving-retrieved XDF signal from the reed diffuser stick (\textit{i}) and PMMA rod (\textit{ii}), along with line profiles across the various retrieved multimodal images in these regions, are shown in Fig.~\ref{fig:FourWood_MultiLineProfile2}. Figure~\ref{fig:FourWood_MultiLineProfile2}b) plots the evolving- and devolving-retrieved XDF values in the interior of the reed diffuser stick (\textit{i}), with a magnified image of the $\tilde{D}_{\textrm{Dev.}}$ in this region shown in Fig.~\ref{fig:FourWood_MultiLineProfile2}a). Although the exact number of fibers traced in this line profile is unknown, more fiber-like structures are discernible in the devolving XDF compared to the evolving XDF. Moreover, the blue trace in Fig.~\ref{fig:FourWood_MultiLineProfile2}b) lacks well-defined troughs and peaks (in terms of visibility), preventing macroscopic fiber identification. In contrast, the orange plot exhibits distinct troughs and peaks, enabling the reed diffuser stick's fiber-like structures to be resolved. The complementarity of the XDF and transmission signals, as well as the two SBXI evolution perspectives, are shown in Fig.~\ref{fig:FourWood_MultiLineProfile2}d), which shows the retrieved signals across the edge of the PMMA rod (\textit{ii}). A magnified view of this interface is shown in Fig.~\ref{fig:FourWood_MultiLineProfile2}c). The evolving- and devolving-retrieved transmission values are approximately identical across this interface with just a 0.05\% difference between the recovered $t_{0_\textrm{Ev.}}$ and $t_{0_\textrm{Dev.}}$ values. The recovered transmission signal at the air--PMMA interface shows no residual phase contrast (i.e., edge enhancement), indicating that phase retrieval has been accurately performed using both the evolving and devolving SBXI Fokker–Planck perspectives. In contrast, the retrieved XDF values across this sharp interface differ significantly between the two perspectives. The devolving-retrieved XDF, represented by the solid orange trace in Fig.~\ref{fig:FourWood_MultiLineProfile2}d), is largely negative at the edge of the PMMA rod and weakly negative inside the rod. Conversely, the evolving-retrieved XDF, shown in solid blue in Fig.~\ref{fig:FourWood_MultiLineProfile2}d), is largely positive at the edge and weakly positive inside the rod. This observation further supports the earlier claim that the devolving model can isolate sharp-edge-induced XDF, separating it into the negative XDF component, whereas the evolving model cannot. Thus the two evolution models recover the XDF signal differently, especially at sharp edges, yet the recovered transmission signals are very consistent.
\begin{figure}[tb!] 
    \centering 
    \includegraphics[width=0.9\linewidth]{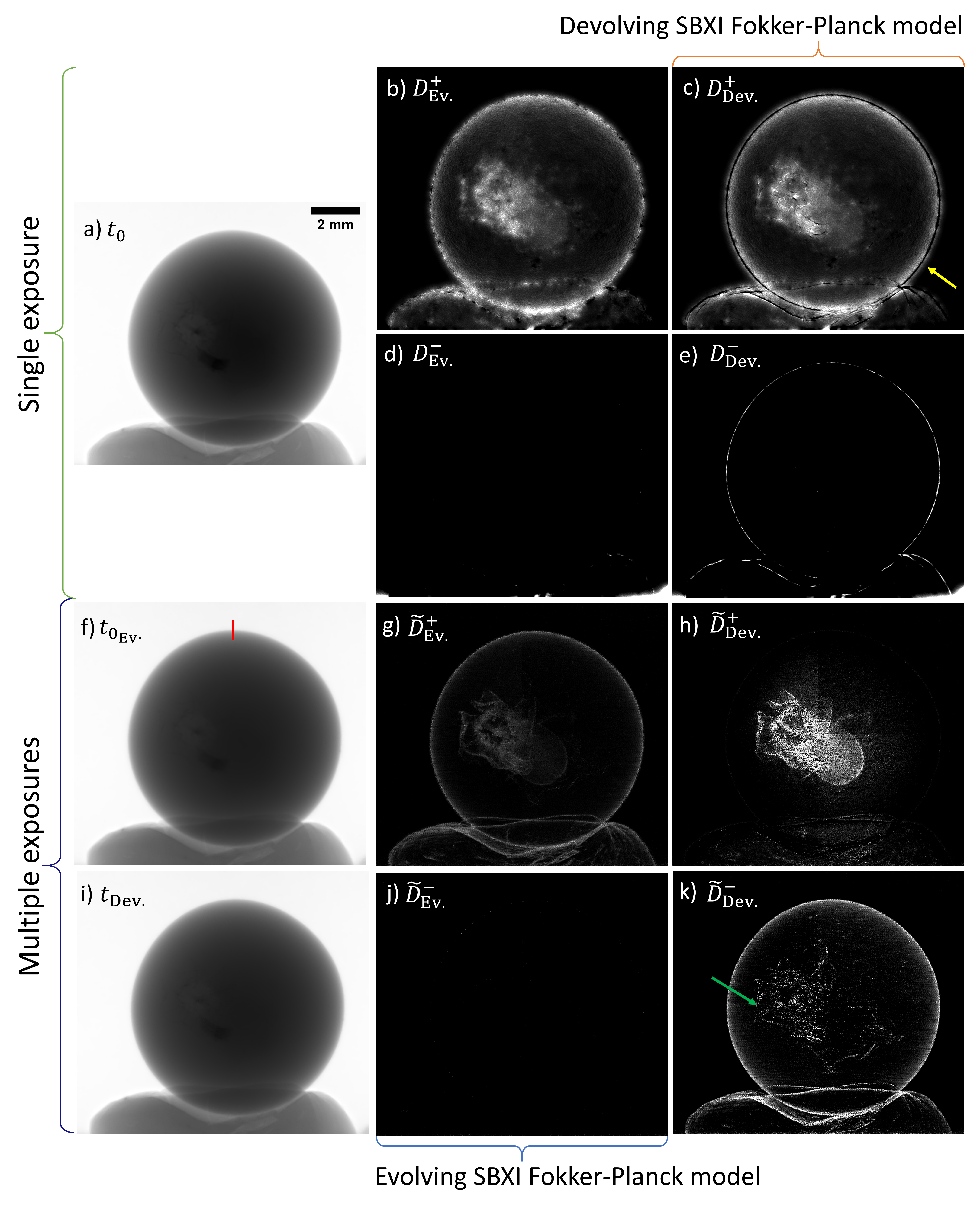}
    \caption{The red currant's recovered transmission and XDF images using evolving and devolving SBXI Fokker--Planck models, as distinguished by the subfigures' subscripts. a)--e) were calculated using the single-exposure approaches and f)--k) with the multiple-exposure approaches (7 sets of SBXI data). The XDF signal is separated into two images showing the magnitude of the positive and negative components. The red annotation in f) indicates the position of intensity profiles shown in Fig.~\ref{fig:RC_CloseLine}d). The yellow arrow in c) marks the bright–dark–bright halo observed in the single-exposure devolving-retrieved XDF at strong phase edges. The green arrow in k) denotes the fibrous network surrounding the red currant pip. Images are shown in linear grayscale with min and max values of [min, max]: a) = f) = i) [0.3, 1], b) = c) = g) = h) [0.1, 8.5]$\times 10^{-11}$, and d) = e) = j) = k) [0.4, 6.1]$\times 10^{-11}$.} 
    \label{fig:RC_Recon2} 
\end{figure}

\subsection{Red currant sample}
The recovered images of the red currant sample, using both the evolving and devolving single- and multiple-exposure methods, are shown in Fig.~\ref{fig:RC_Recon2}. As seen for the four-rod sample, useful information about sample-imposed X-ray diffusion under the evolving SBXI Fokker--Planck perspective is only contained within the positive XDF signal; see Figs.~\ref{fig:RC_Recon2}b) and g) which show the positive components of the evolving-retrieved XDF signal using the single- and multiple-exposure approach, respectively, and compare these with Figs.~\ref{fig:RC_Recon2}d) and j) which show the corresponding negative components. The red currant's devolving-retrieved XDF, however, contains both positive and negative values: positive values appear in structure-dense regions, such as the pip, while negative XDF signals are observed at sharp edges, like the red currant--air interface. This is observed in both the single- and multiple-exposure reconstructions but more clearly in the latter. Single-exposure SBXI algorithms can have difficulty interpreting intensity variations due to propagation-based phase contrast or attenuation that are similar in size to the speckles. For example, in the devolving single-exposure reconstruction, there is a halo-like artefact around the red currant, indicated by the yellow arrow in Fig.~\ref{fig:RC_Recon2}c). 
Such difficulties associated with single-exposure algorithms may be mitigated when using multiple SBXI speckle positions, as features in the sample are probed with different speckles and changes in the local mean intensity across multiple pairs of SBXI data can be isolated. The red currant's sharp edge is well-reconstructed in the XDF using the multiple-exposure retrieval algorithms, as shown in Figs.~\ref{fig:RC_Recon2}g) and k). 

\begin{figure}[tb!] 
    \centering 
    \includegraphics[width=\linewidth, trim=5 5 5 5,clip]{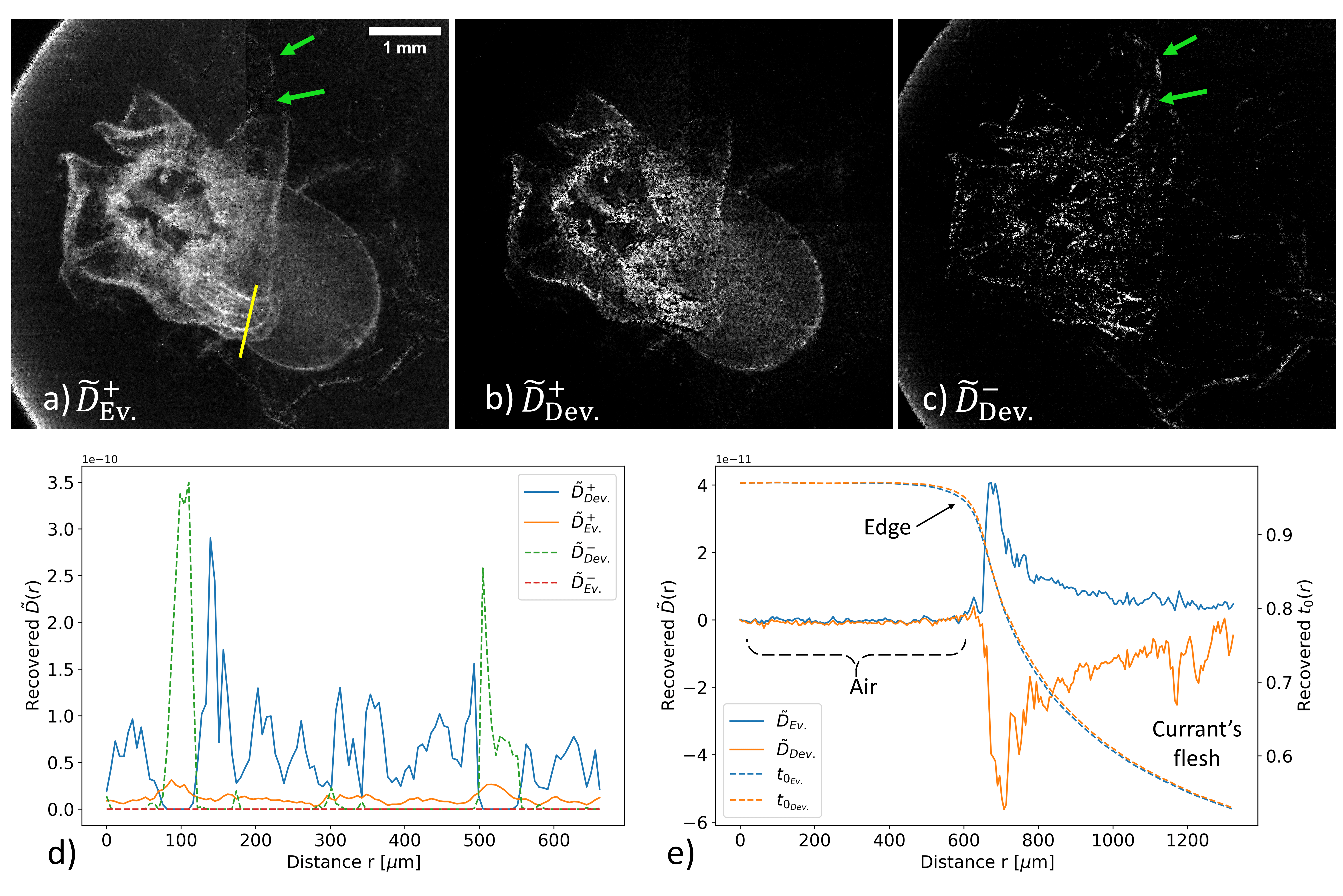}
    \caption{a), b), and c) are magnified images of the XDF signal extracted from the red currant's pip: a) positive XDF component using the evolving Fokker--Planck multiple-exposure method. As $\tilde{D}_{\textrm{Ev.}}^{-}$ is negligible/weak, as shown by Figs.~\ref{fig:RC_Recon2}j), it is omitted from this figure. b) and c) show the magnitudes of the positive and negative XDF components, respectively, using the devolving multiple-exposure approach. d) plots the magnitudes of the XDF components along the yellow trace in a). e) plots the reconstructed XDF and transmission across the air--currant interface, indicated by the red line in Fig.~\ref{fig:RC_Recon2}f), using both models. Images are shown in linear grayscale with min and max values of [min, max]: a) [0.0, 2.7]$\times10^{-11}$, b) [0.8, 17.8]$\times10^{-11}$, and c) [0.8, 8.5]$\times10^{-11}$.
    } 
    \label{fig:RC_CloseLine} 
\end{figure}

A red currant's pip contains structures across a range of length scales. Consequently, an incident X-ray beam may be diffused via the mechanisms of SAXS and multiple refraction by structures contained within the pip. Surrounding the pip is a fibrous network, indicated by the green arrow in Fig.~\ref{fig:RC_Recon2}k). This network is large relative to the spatial resolution of the imaging system, making the fibers' edges resolvable in the transmission image (when the maximum and minimum gray levels are adjusted appropriately). These two features, the highly structured pip and the sharp-edged fibrous network, are clearly separated in the devolving-retrieved XDF. Figure~\ref{fig:RC_Recon2}h) shows the multiple-exposure $\tilde{D}_{\textrm{Dev.}}^{+}$ signal, where only the red currant's pip produces a retrievable positive XDF. Figure~\ref{fig:RC_Recon2}k) shows the negative component of the devolving-retrieved XDF, where the edges of both the red currant and the fibrous network around the pip are isolated. This separation of edge and microstructure XDF contrast is also visible in images of the red currant retrieved using a single SBXI speckle position, Figs.~\ref{fig:RC_Recon2}c) and e). However, fine scattering features, such as the edge of the red currant and the pip, are more distinctly isolated using multiple-exposure methods, Figs.~\ref{fig:RC_Recon2}h) and k).

Figure~\ref{fig:RC_CloseLine} presents magnified images of the red currant pip's XDF retrieved using the multiple-exposure approaches, as well as intensity plots across two traces of the red currant's signals. This figure further evidences the ability of the devolving perspective-based algorithms to distinguish the difference between the two distinct XDF mechanisms. Additionally, finer structures are discernible in the devolving-retrieved XDF that are not visible in the evolving-retrieved XDF. Several pip fibers and surrounding features are present in Fig.~\ref{fig:RC_CloseLine}c) ($\tilde{D}_{\textrm{Dev.}}^{-}$), and many of these cannot be discerned in Fig.~\ref{fig:RC_CloseLine}a) ($\tilde{D}_{\textrm{Ev.}}^{+}$); two examples are indicated by the green arrows in Figs.~\ref{fig:RC_CloseLine}a) and c). This is further proven by the line profiles shown in Fig.~\ref{fig:RC_CloseLine}d). The line profile across $\tilde{D}_{\textrm{Dev.}}^{-}$ (green-dashed plot) has two distinct peaks at the location of two fiber-like features. Such peaks are also visible in the $\tilde{D}_{\textrm{Ev.}}^{+}$ signal (solid-orange plot); however, their magnitude is much smaller, indicating reduced visibility of these features in the evolving-retrieved XDF compared to the devolving-retrieved XDF signal. The blue-solid plot shows the $\tilde{D}_{\textrm{Dev.}}^{+}$ values where there are regions of dense microstructure. Finally, Fig.~\ref{fig:RC_CloseLine}e) shows the reconstructed XDF and transmission values on a line profile taken across the air--red currant flesh interface, denoted by the red line in Fig.~\ref{fig:RC_Recon2}f). This plot has similar characteristics to that shown for the four-rod sample in Fig.~\ref{fig:FourWood_MultiLineProfile2}d): 1) The evolving- and devolving-retrieved transmission signals are very consistent across a sharp edge, with just a 0.3\% difference between the transmission values in Fig.~\ref{fig:RC_CloseLine}e). 2) The phase retrieval was performed accurately using both Fokker–Planck evolution models, with sharp edges being reconstructed sharply and without any edge enhancement nor over-smoothing. 3) A sharp edge's XDF signal is positive under the evolving perspective and negative under the devolving perspective. Notably, the signal-to-noise ratio (SNR)\footnote{The signal-to-noise ratio was calculated by taking the mean intensity of a local region in the center of the red currant, where XDF is generated, divided by the standard deviation of a region in air outside the red currant.} in the devolving-retrieved XDF is 6.4 times greater than in the evolving-retrieved XDF; SNR$_\textrm{Ev.\:XDF}$ = 8.09 and SNR$_\textrm{Dev.\:XDF}$ = 51.72.

\section{Discussion}
The major XDF-contrast mechanisms in Fokker--Planck-based SBXI algorithms are diffusion related to position-dependent SAXS, multiple refraction on large structures, and diffuse X-ray scatter at sharp edges \cite{pavlov2020x}. In the experimental results section, we have seen: (1) The recovered XDF signals from the evolving and devolving SBXI Fokker--Planck perspectives are different, but the retrieved transmission signals are virtually identical. (2) From the evolving perspective, all of the useful sample diffusion information is contained within the positive XDF signal. (3) The devolving-retrieved XDF signal contains significant positive and negative values, each originating from different types of sample structures and hence contrast mechanisms. Specifically, positive values are found in structure-dense regions, which induce local SAXS and/or multiple refraction, while negative XDF values are recovered from sharp resolvable edges in a sample. The separation of microstructure and sharp edges in the XDF image using devolving Fokker--Planck algorithms provides sample information separation that is not directly accessible from the evolving algorithms. To our knowledge, the theory does not currently exist to explain the recovered negative diffusion coefficient when using the devolving Fokker--Planck equation for paraxial X-ray imaging. While such a theoretical development is beyond the scope of the present paper, in the text below we evaluate the relevant research literature and suggest some avenues for future work. 

\subsection{Kolmogorov equations}
What has been called here the \textit{evolving} Fokker--Planck equation is a type of forward Kolmogorov equation. The forward Kolmogorov equation is one of the two `Kolmogorov equations', the other being the backward Kolmogorov equation. In general, the Kolmogorov equations describe the evolution of a probability density function for a given stochastic process \cite{risken1989fokkerplanck,Oksendal2007,Bjork2009,pavliotis2014stochastic,bogachev2015fokker}. Considering a process' evolution from position $x_0$ at time $t_0$ to $x$ at a later time $t$ gives the forward Kolmogorov equation. The backward Kolmogorov equation considers the same process but in the opposite temporal direction. The backward Kolmogorov equation is useful when considering an evolution problem in which the final state of the system is known, and the probability density function at an earlier point in time is of interest \cite{risken1989fokkerplanck, pavliotis2014stochastic}. When Paganin and Morgan \cite{paganin2019x} derived the paraxial-optics X-ray Fokker--Planck equation from the general Fokker--Planck equation, the time dependence was replaced with the propagation distance dependence along the optical axis, $z$. In SBXI, an image is captured some distance $z = \Delta$ downstream of the sample, and the solution to the multimodal inverse problem aims to retrieve sample characteristics by reconstructing the wavefield at $z=0$, immediately after the sample. Potentially, this simple analogy to the general forward and backward Kolmogorov equations suggests that, indeed, the devolving Fokker--Planck equation (presented originally in Beltran \textit{et al.} \cite{beltran2023} and also in this paper) should be used when solving the associated SBXI multimodal inverse problem.

\subsection{Tracking speckles forwards and backwards}

A similar change in the `direction of evolution' consideration was recently suggested in Morgan \textit{et al.} \cite{morgan2020ptychographic}, in the context of the UMPA speckle-tracking algorithm \cite{zdora2017x}, which was being used within speckle ptychography. Morgan \textit{et al.} proposed that this new `reverse UMPA' would provide better results. This was then explored in De Marco \textit{et al.} \cite{de2023high}. Essentially, the originally published UMPA algorithm assigns sample-imposed speckle modulations to a position in the sample-plus-speckle image. The revised variant \cite{morgan2020ptychographic} assigns the sample-induced speckle modifications to the position of the speckle in the reference-speckle image plane -- which is closer to the point of interaction, and thus, it potentially makes more sense to use this model. The difference in the position to which the speckle modifications are assigned in the two variants of UMPA would only result in significantly different retrieved images if the speckle modifications are sufficiently large, such as at sharp edges\footnote{This significant difference between the two evolution models at sharp sample edges was identified during personal communications with Luca Fardin, from a theoretical image-registration perspective.}. This might help explain why the retrieved images in the evolving and devolving Fokker--Planck models only diverge at sharp edges. Interestingly, both UMPA-algorithm variants struggled to track speckles at sample edges with strong propagation-based phase-contrast fringes, resulting in artefacts resembling bright/dark speckles in the phase (see e.g. Fig.~6 in Ref.~\citenum{de2023high}) and XDF images. UMPA and MIST are both multiple-exposure SBXI approaches. However, unlike UMPA, both evolution variants of the multiple-exposure MIST algorithm successfully recovered an artefact-free XDF signal—without bright/dark speckle structures—at sharp edges with strong propagation-based phase contrast in the experimental SBXI data. The difference in UMPA's and MIST's retrieval abilities at sharp edges may stem from the different manner in which each algorithm models phase effects within its theoretical framework. Specifically, UMPA does not model propagation-based phase-contrast fringes from sample edges. In contrast, the algorithms explored here do so via the paraxial-optics Fokker--Planck equation's lensing term. The difference between UMPA and MIST's retrieved edge signals may also stem from the algorithms' distinct computational procedures. Specifically, UMPA uses a finite-sized analysis window in both the reference and sample speckle images to locally search for speckle modifications. This means that the UMPA algorithm has difficulty quantifying reference speckle modifications when much of the analysis window is significantly distorted, such as by a strong phase edge. In contrast, MIST operates on a whole-image level, so defined analysis windows do not constrain its computational approach.

\subsection{Edge signals}
The XDF retrieval at sample edges can be modelled using diffusion operators. In this section of the discussion, we seek to better understand the behaviour of the two Fokker--Planck retrieval methods by discussing these operators and associated stability. Under the evolving perspective, speckles are blurred in the presence of an X-ray-diffusing material, as represented by the third term on the right-hand side of Eq.~\ref{eqn:FPphase_Evolve}. Conversely, speckles are sharpened due to the removal of an X-ray-diffusing material in the devolving model, as shown by the third term on the right-hand side of Eq.~\ref{eqn:FPphase_Devolve}. 
\begin{figure}[tb!]
    \centering 
    \includegraphics[width=\linewidth, trim=5 5 5 5,clip]{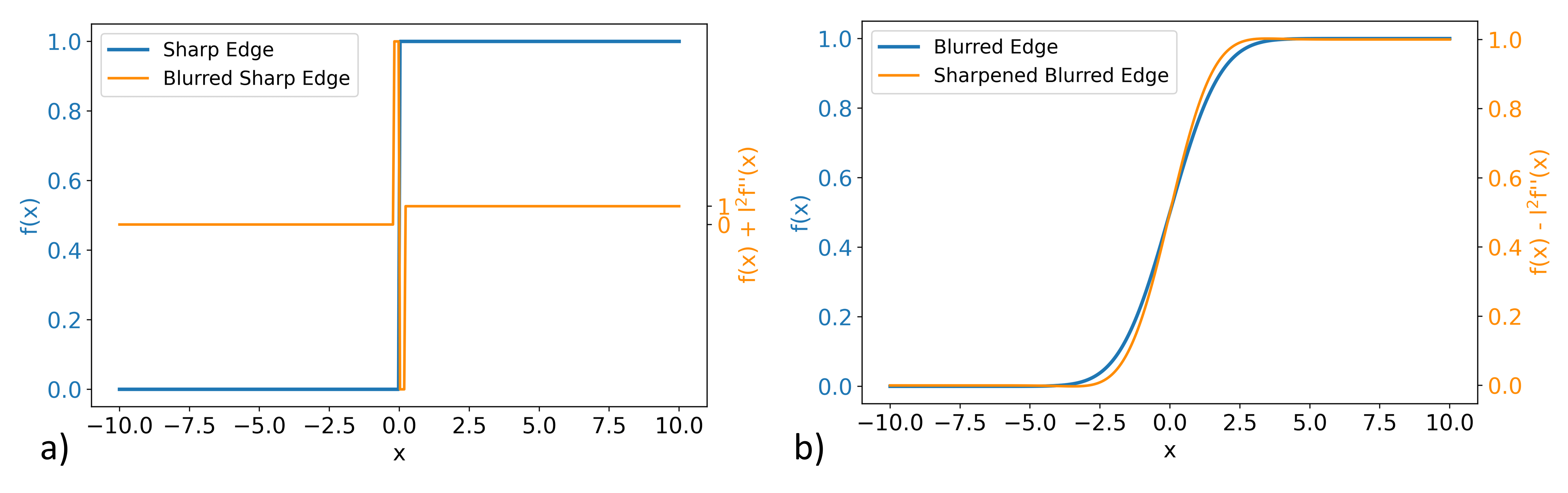} 
    \caption{Mathematical representations of the operators in the evolving and devolving SBXI Fokker--Planck models acting on an edge. a) Evolving model: The blurring operator $\hat{g}_\textrm{Blur}(x)=[1+l^2\partial^2/\partial x^2]$ acts on a sharp edge, defined by the Heaviside function (blue plot). The result is shown in orange, which demonstrates the breakdown of the evolving Fokker--Planck diffusion model at sharp edges. b) Devolving model: The sharpening operator $\hat{g}_\textrm{Sharpen}(x) = [1-l^2\partial^2/\partial x^2]$ is applied to a blurred edge. The blurred edge, shown in blue, is modelled by an error function $erf(x)$. The result of applying the sharpening operator to this function is shown by the orange plot. An arbitrary blur width of $l^2=0.4$ was used in both operators to generate both orange plots in this figure. }
    \label{fig:StabilityOfOperations} 
\end{figure}
First, we will analyze how the evolving model addresses sharp edges within its framework. A one-dimensional blurring operator similar to the Fokker--Planck model has the approximate form $\hat{g}_\textrm{Blur}(x) = [1+l^2\partial^2/\partial x^2]$, where $l$ is the blur width which characterizes the degree of blurring. To consider edge-induced diffusion, the evolving model applies this blurring operator to a sharp edge. A Heaviside function, for example, can model the sharp edge: see the blue curve in Fig.~\ref{fig:StabilityOfOperations}a). Applying the blurring operator yields the orange plot in Fig.~\ref{fig:StabilityOfOperations}a). This operator is expected to smooth or blur the Heaviside function over a finite width; however, we instead observe two opposite, infinitely tall spikes or impulses on either side of the boundary. The evolving Fokker--Planck equation uses a similar operator to model diffusion, which is unstable at sharp edges, as demonstrated in Fig.~\ref{fig:StabilityOfOperations}a).

Now, in the devolving Fokker--Planck model, the sharpening operator has the approximate form $\hat{g}_\textrm{Sharpen}(x) = [1-l^2\partial^2/\partial x^2]$ \footnote{Note that $\hat{g}_\textrm{Blur}(x)$ and $\hat{g}_\textrm{Sharpen}(x)$ are inverses to a third-order approximation.}. The devolving perspective considers edge-induced X-ray diffusion via a blurred edge getting sharpened. Thus, the sharpening operator can be applied to an error function $erf(x)$ (which serves as a one-dimensional model for a blurred edge) to investigate the stability of the devolving SBXI Fokker--Planck equation at edges. An error function is plotted in blue in Fig.~\ref{fig:StabilityOfOperations}b) and the orange plot in this figure shows the result of applying $\hat{g}_\textrm{Sharpen}(x)$ to this blurred edge. Indeed, the orange plot is a sharpened version of the blue plot. By analyzing the general one-dimensional form of diffusive optical flow-operators from both Fokker--Planck perspectives and applying them to edges, we see that the devolving model is stable at sample edges, while the evolving model is not.
\newpage
\subsection{Physical meaning of Fokker--Planck diffusion}
The effective diffusion coefficient, $D$, quantifies the extent to which the sample diffuses the incident X-ray wavefield. This coefficient is related to the opening angle of the SAXS fans emanating from the exit surface of the sample via $D = (F \theta^2)/2$ \cite{paganin2019x}. Positive $D$ values indicate that X-rays are locally diffused outwards in random directions. Negative $D$ values, as observed in the devolving-retrieved XDF in this work, represent local sharpening. A potential justification for the negative $D$ signal recovered at sharp edges, is to consider the phenomenon of Young-Maggi-Rubinowicz boundary waves \cite{BornWolf1999,borghi2015uniform} or diffracted rays (from a ray-optics perspective) \cite{keller1962geometrical}. Born and Wolf \cite{BornWolf1999} describe the Young-Maggi-Rubinowicz boundary-wave mechanism as being `the appearance of a diffracting edge being luminous when observed within the geometric shadow'. The physical existence of these Young-Maggi-Rubinowicz boundary waves has been experimentally validated \cite{kumar2007direct}. Therefore, it is reasonable to propose that such phenomena are revealing themselves in SBXI experiments and, hence, the XDF images we have presented in this manuscript. From the boundary-waves viewpoint, X-rays will be sharpened for a small number of pixels directly in line with the edges of the sample from the detector's perspective. This sharpening is represented by a purely imaginary blur width \cite{croughan2023directional}, which is analogous to recovering a negative $D$ value. The mention of negative diffusion coefficients in a Fokker--Planck context is sparse in the research literature. However, there are some early reports of negative diffusion within the quantum optics field \cite{gronchi1979fokker, yuen1986langevin,tan1987solution}.

\subsection{Phase and dark-field cross-talk at edges}
Another possibility is that the negative diffusion coefficient, recovered when using the devolving SBXI Fokker--Planck perspective, results from sharp edge phase `contamination'. The work of Morgan and Paganin \cite{morgan2019applying} explored the application of the paraxial-optics Fokker--Planck equation to a grating-interferometry technique. The work aimed to investigate cross-talks between X-ray transmission, phase, and XDF signals. Although the work was derived for the case of a sinusoidal reference pattern, it can be applied in the context of SBXI, as a reference speckle pattern can be decomposed into appropriate sinusoidal components which sum through Fourier analysis to produce the random speckle pattern. A related result of Morgan and Paganin \cite{morgan2019applying} is that diffuse scattering at the boundary between two distinctly different scattering regions and/or a strong phase edge can increase a reference pattern's visibility (corresponding to a negative effective diffusion coefficient, as used here). The latter effect suggests potential phase and XDF cross-talk within the Fokker–Planck model; however, this seems unlikely or remains unclear, given the results in Figs.~\ref{fig:FourWood_MultiLineProfile2}d) and \ref{fig:RC_CloseLine}e). Figures~\ref{fig:FourWood_MultiLineProfile2}d) and \ref{fig:RC_CloseLine}e) show that the edge enhancement due to propagation-based phase contrast in the raw SBXI intensity data is accurately managed by the derived MIST algorithms; the phase edges are sharply recovered, with no residual phase contrast at edges in the retrieved transmission images. Interestingly, there is only the XDF signal at edges, suggesting that the XDF observed at these locations arises from a fundamentally different mechanism than propagation-based phase contrast. It should be noted that some XDF literature, for different experimental techniques, claim that phase-edges that are comparable in size to the reference illumination features directly contaminate their recovered XDF signal, in the sense that phase effects can be directly divided out of the raw intensity images before performing the single-exposure XDF retrieval, see Fig.~3 in Croughan \textit{et al.} \cite{croughan2024correcting}. This correction was applied to some SBXI data (not those used in this work), and this was then used within our developed MIST algorithms, however it was unsuccessful as edge-induced XDF was still retrieved. This suggests further that the underlying contrast mechanism at sharp edges in the transmission and XDF images are different from each other and those seen in other experimental techniques. Contribution to edge signals may arise during the retrieval process from differences between experimental data and the propagation-based phase-contrast fringe description provided by the TIE/Fokker--Planck equation, which uses a Laplacian to describe a near-field fringe with no point-spread-function blurring. The next questions are, `Is the boundary-wave phenomenon the underlying mechanism of the observed local sharpening?'  and `Why can just the devolving Fokker--Planck perspective observe this sharpening and separate it from local blurring?'. These questions are left for future research work. 

\subsection{Edge X-ray dark-field in other experimental techniques}
A notable and relevant study is that of Gureyev \textit{et al.} \cite{gureyev2020dark}, who similarly reported the occurrence of negative XDF signals at sample edges when using their free-space propagation XDF retrieval algorithm. This is interesting as the algorithm is fundamentally different (derived from different theoretical considerations) from the devolving Fokker--Planck algorithms presented here. Edge-induced XDF effects have also been explored in grating-interferometry imaging techniques \cite{yashiro2015effects}. Specifically, Yashiro and Momose \cite{yashiro2015effects} theoretically and experimentally demonstrated that sharp edges generate visibility contrast, i.e., generate an XDF signal. However, the ability to separate this XDF contrast mechanism from microstructure scatter contribution was not addressed in this work. It was noted, however, that it might be possible to separate the different visibility-contrast mechanisms in grating-interferometry by varying experimental parameters such as grating pitch, the distance between gratings, and wavelength, and then analyzing how the XDF changes for particular sample features. In our work, we have successfully performed the described discrimination by employing the devolving Fokker--Planck equation for paraxial X-ray imaging.
In grating interferometry and other experimental techniques, strong edge XDF, which can contaminate regions of interest within the sample, is often avoided by embedding the sample prior to imaging in a material with a similar refractive index \cite{loo2001new}. The approach presented here provides a computational approach to separate this edge XDF, eliminating the need for additional sample preparation. Furthermore, sample embedding is useful in computed tomography to prevent strong edge signals from causing streaking artefacts in reconstructed axial slices. The edge XDF separation demonstrated here could help suppress such streaking artefacts. Accordingly, in future work, it would be interesting to investigate the relationship between these different XDF retrieval approaches.

\section{Conclusion}
This work investigated the two perspectives, \textit{evolving} and \textit{devolving}, of the Fokker--Planck equation for paraxial X-ray imaging in the context of speckle-based X-ray imaging (SBXI). The evolving perspective considers how reference X-ray speckles are altered when a sample is placed into the X-ray beam. The devolving perspective tracks how intensity modulations introduced by both the sample and the speckle-membrane are changed when the sample is removed from the SBXI system. Although these perspectives are physically equivalent, they are potentially mathematically inequivalent. Hence, this study aimed to experimentally investigate the points of agreement and differences between the corresponding distinct approaches to the multimodal inverse problem of the two Fokker--Planck equation perspectives. A single- and multiple-exposure multimodal---transmission, phase, and X-ray dark-field (XDF)---algorithm was given for each perspective. The derived algorithms were then applied to two sets of experimental SBXI data, and the differences in the recovered images were examined. The key finding of this work is that the evolving and devolving perspectives attribute the microstructure- and edge-induced XDF contrast to different signs of the recovered effective diffusion coefficient. Moreover, a sample's evolving-retrieved XDF signal is entirely positive; no useful information is contained within the negative component. In contrast, the devolving-retrieved XDF signal is positive in microstructure-dense regions and negative at sharp edges. It follows that the devolving Fokker--Planck algorithms can separate two physically different diffuse-scatter mechanisms, diffuse scatter due to spatially random microstructure and diffuse scatter from sharp edges. Our results suggest that the devolving variant of the Fokker--Planck equation is more suitable for solving the multimodal SBXI inverse problem compared to the evolving equation. Future research work, guided by the experimental results presented in this paper, will involve theoretically understanding X-ray diffusion at sharp edges in the context of Fokker--Planck-type diffusion, how this edge diffusion reveals itself in SBXI data, and why the two different perspectives of the SBXI Fokker--Planck equation treat this mechanism differently. The results of this work may be useful in eliminating the need to embed objects that generate strong phase contrast in materials with similar refractive indices \cite{loo2001new}. The ability to separate edge-induced XDF may also help suppress streak artefacts in phase and XDF computed tomography reconstructions, which often result from strong edge scattering. Finally, the results and discussion in this work may provide insights into the potential coupling of phase and diffusion effects at sharp edges.
\section*{Acknowledgments}
We gratefully acknowledge useful discussions with Luca Fardin. The authors thank Sebastien Berujon, Eric Ziegler, and Emmanuel Brun for collecting and sharing the red currant SBXI data collected at European Synchrotron Radiation Facility (ESRF), originally published in Berujon \textit{et al.} \cite{berujon2016x}. The authors are grateful for the help provided by the beamline scientists Andrew Stevenson and Benedicta Arhatari at the MicroCT beamline at the Australian Synchrotron, where the images in Figs.~3 and 4 were captured under proposal 19633. We acknowledge the University of Canterbury for awarding a doctoral scholarship to Samantha J. Alloo, and the Australian Institute of Nuclear Science and Engineering (AINSE Ltd.), as this research was supported by a Postgraduate Research Award (PGRA). Kaye S. Morgan acknowledges support from the Australian Research Council (FT18010037 and DP230101327). Jannis N. Ahlers and Michelle K. Croughan acknowledge support from an Australian Government Research Training Program (RTP) Scholarship.


\bibliographystyle{abbrv}  
\bibliography{references}  

\begin{thebibliography}{10}

\bibitem{ahlers2024x}
J.~N. Ahlers, K.~M. Pavlov, M.~J. Kitchen, and K.~S. Morgan.
\newblock X-ray dark-field via spectral propagation-based imaging.
\newblock {\em Optica}, 11(8):1182--1191, 2024.

\bibitem{reconstruction_Github}
S.~J. Alloo.
\newblock {Reconstructions, Data, and Script for Evolving--Devolving Speckle-Based Fokker--Planck Comparison}, 2024.
\newblock \href{https://github.com/samanthaalloo/EvVSDev_SBXIFokkerPlanck}{Available on GitHub}, accessed on 23 October 2024.

\bibitem{alloo2023m}
S.~J. Alloo, K.~S. Morgan, D.~M. Paganin, and K.~M. Pavlov.
\newblock Multimodal intrinsic speckle-tracking ({MIST}) to extract images of rapidly-varying diffuse x-ray dark-field.
\newblock {\em Scientific Reports}, 13(1):5424, 2023.

\bibitem{alloo2021speckle}
S.~J. Alloo, D.~M. Paganin, K.~S. Morgan, M.~J. Kitchen, A.~W. Stevenson, S.~C. Mayo, H.~T. Li, B.~Kennedy, A.~Maksimenko, J.~Bowden, and K.~M. Pavlov.
\newblock Speckle-based x-ray dark-field tomography of an attenuating object.
\newblock In {\em Developments in X-Ray Tomography XIII}, volume 11840, pages 78--90. SPIE, 2021.

\bibitem{alloo2022dark}
S.~J. Alloo, D.~M. Paganin, K.~S. Morgan, M.~J. Kitchen, A.~W. Stevenson, S.~C. Mayo, H.~T. Li, B.~M. Kennedy, A.~Maksimenko, J.~C. Bowden, J.~Bowden, and K.~M. Pavlov.
\newblock Dark-field tomography of an attenuating object using intrinsic x-ray speckle tracking.
\newblock {\em Journal of Medical Imaging}, 9(3):031502, 2022.

\bibitem{aminzadeh2022imaging}
A.~Aminzadeh, B.~D. Arhatari, A.~Maksimenko, C.~J. Hall, D.~Hausermann, A.~G. Peele, J.~Fox, B.~Kumar, Z.~Prodanovic, M.~Dimmock, D.~Lockie, K.~M. Pavlov, Y.~I. Nesterests, D.~Thompson, S.~C. Mayo, D.~M. Paganin, S.~T. Taba, S.~Lewis, P.~C. Brennan, H.~M. Quiney, and T.~E. Gureyev.
\newblock Imaging breast microcalcifications using dark-field signal in propagation-based phase-contrast tomography.
\newblock {\em IEEE Transactions on Medical Imaging}, 41(11):2980--2990, 2022.

\bibitem{arhatari2023micro}
B.~D. Arhatari, A.~W. Stevenson, D.~Thompson, A.~Walsh, T.~Fiala, G.~Ruben, N.~Afshar, S.~Ozbilgen, T.~Feng, S.~Mudie, and P.~Tissa.
\newblock {Micro-Computed Tomography beamline of the Australian Synchrotron: Micron-size spatial resolution x-ray imaging}.
\newblock {\em Applied Sciences}, 13(3):1317, 2023.

\bibitem{beltran2023}
M.~A. Beltran, D.~M. Paganin, M.~K. Croughan, and K.~S. Morgan.
\newblock Fast implicit diffusive dark-field retrieval for single-exposure, single-mask x-ray imaging.
\newblock {\em Optica}, 10(4):422--429, April 2023.

\bibitem{Bennett2010}
E.~E. Bennett, R.~Kopace, A.~F. Stein, and H.~Wen.
\newblock A grating-based single-shot x-ray phase contrast and diffraction method for in vivo imaging.
\newblock {\em Medical Physics}, 37(11):6047--6054, 2010.

\bibitem{berujon2012x}
S.~B{\'e}rujon, H.~Wang, and K.~Sawhney.
\newblock X-ray multimodal imaging using a random-phase object.
\newblock {\em Physical Review A}, 86(6):063813, 2012.

\bibitem{berujon2016x}
S.~B{\'e}rujon and E.~Ziegler.
\newblock X-ray multimodal tomography using speckle-vector tracking.
\newblock {\em Physical Review Applied}, 5(4):044014, 2016.

\bibitem{berujon2017near}
S.~B{\'e}rujon and E.~Ziegler.
\newblock Near-field speckle-scanning-based x-ray tomography.
\newblock {\em Physical Review A}, 95(6):063822, 2017.

\bibitem{berujon2012two}
S.~B{\'e}rujon, E.~Ziegler, R.~Cerbino, and L.~Peverini.
\newblock Two-dimensional x-ray beam phase sensing.
\newblock {\em Physical Review Letters}, 108(15):158102, 2012.

\bibitem{Bjork2009}
T.~Bj{\"o}rk.
\newblock {\em {Arbitrage Theory in Continuous Time}}.
\newblock Oxford University Press Inc, New York, 3rd edition, 2009.

\bibitem{bogachev2015fokker}
V.~I. Bogachev, N.~V. Krylov, M.~R{\"o}ckner, and S.~V. Shaposhnikov.
\newblock {\em {Fokker--Planck--Kolmogorov Equations}}.
\newblock American Mathematical Society, 2015.

\bibitem{borghi2015uniform}
R.~Borghi.
\newblock Uniform asymptotics of paraxial boundary diffraction waves.
\newblock {\em Journal of the Optical Society of America A}, 32(4):685--696, 2015.

\bibitem{BornWolf1999}
M.~Born and E.~Wolf.
\newblock {\em {Principles of Optics}}.
\newblock Cambridge University Press, Cambridge, 7th edition, 1999.

\bibitem{carmichael2013statistical}
H.~J. Carmichael.
\newblock {\em {Statistical Methods in Quantum Optics 1: Master Equations and Fokker--Planck Equations}}.
\newblock Springer Science \& Business Media, 1st edition, 1998.

\bibitem{chavanis2003generalized}
P.-H. Chavanis.
\newblock {Generalized thermodynamics and Fokker--Planck equations: Applications to stellar dynamics and two-dimensional turbulence}.
\newblock {\em Physical Review E}, 68(3 Pt 2):036108, 2003.

\bibitem{cooper1971compton}
G.~Cooper.
\newblock Compton {{Fokker--Planck}} equation for hot plasmas.
\newblock {\em Physical Review D}, 3(10):2312--2316, 1971.

\bibitem{croughan2023directional}
M.~K. Croughan, Y.~Y. How, A.~Pennings, and K.~S. Morgan.
\newblock Directional dark-field retrieval with single-grid x-ray imaging.
\newblock {\em Optics Express}, 31(7):11578--11597, 2023.

\bibitem{croughan2024correcting}
M.~K. Croughan, D.~M. Paganin, S.~J. Alloo, J.~N. Ahlers, Y.~Y. How, S.~A. Harker, and K.~S. Morgan.
\newblock Correcting directional dark field x-ray imaging artefacts using position dependent image deblurring and attenuation removal.
\newblock {\em Scientific Reports}, 14(1):17807, 2024.

\bibitem{de2023high}
F.~De~Marco, S.~Savatovi{\'c}, R.~Smith, V.~Di~Trapani, M.~Margini, G.~Lautizi, and P.~Thibault.
\newblock {High-speed processing of X-ray wavefront marking data with the Unified Modulated Pattern Analysis (UMPA) model}.
\newblock {\em Optics Express}, 31(1):635--650, 2023.

\bibitem{drummond1980generalised}
P.~D. Drummond and C.~W. Gardiner.
\newblock {Generalised P-representations in quantum optics}.
\newblock {\em Journal of Physics A: Mathematical and General}, 13(7):2353--2368, 1980.

\bibitem{Endrizzi2014}
M.~Endrizzi, P.~C. Diemoz, T.~P. Millard, J.~L. Jones, R.~D. Speller, I.~K. Robinson, and A.~Olivo.
\newblock Hard x-ray dark-field imaging with incoherent sample illumination.
\newblock {\em Applied Physics Letters}, 104(2):024106, 2014.

\bibitem{endrizzi2015edge}
M.~Endrizzi, B.~I.~S. Murat, P.~Fromme, and A.~Olivo.
\newblock Edge-illumination x-ray dark-field imaging for visualising defects in composite structures.
\newblock {\em Composite Structures}, 134:895--899, 2015.

\bibitem{gassert2023dark}
F.~T. Gassert, T.~Urban, A.~Kufner, M.~Frank, G.~C. Feuerriegel, T.~Baum, M.~R. Makowski, C.~Braun, D.~Pfeiffer, B.~J. Schwaiger, F.~Pfeiffer, and A.~S. Gersing.
\newblock {Dark-field X-ray imaging for the assessment of osteoporosis in human lumbar spine specimens}.
\newblock {\em Frontiers in Physiology}, 14:1217007, 2023.

\bibitem{goodman2020speckle}
J.~W. Goodman.
\newblock {\em {Speckle Phenomena in Optics: Theory and Applications}}.
\newblock SPIE Press, Bellingham, Washington, 2nd edition, 2020.

\bibitem{gorji2012kinetic}
H.~Gorji and P.~Jenny.
\newblock A kinetic model for gas mixtures based on a {{Fokker--Planck}} equation.
\newblock In {\em Journal of Physics: Conference Series}, volume 362, pages 12042--12047. IOP Publishing, 2012.

\bibitem{gronchi1979fokker}
M.~Gronchi and L.~A. Lugiato.
\newblock Fokker-{P}lanck equation for optical bistability.
\newblock {\em Lettere al Nuovo Cimento}, 23(16):593--598, 1979.

\bibitem{gureyev2020dark}
T.~E. Gureyev, D.~M. Paganin, B.~Arhatari, S.~T. Taba, S.~Lewis, P.~C. Brennan, and H.~M. Quiney.
\newblock Dark-field signal extraction in propagation-based phase-contrast imaging.
\newblock {\em Physics in Medicine \& Biology}, 65(21):215029, 2020.

\bibitem{he2024nondestructive}
J.~He, L.~Van~Doorselaer, A.~Tempelaere, J.~Vignero, W.~Saeys, H.~Bosmans, P.~Verboven, and B.~Nicolai.
\newblock {Nondestructive internal disorders detection of ‘Braeburn’apple fruit by X-ray dark-field imaging and machine learning}.
\newblock {\em Postharvest Biology and Technology}, 214:112981, 2024.

\bibitem{how2022quantifying}
Y.~Y. How and K.~S. Morgan.
\newblock Quantifying the x-ray dark-field signal in single-grid imaging.
\newblock {\em Optics Express}, 30(7):10899--10918, 2022.

\bibitem{keller1962geometrical}
J.~B. Keller.
\newblock Geometrical theory of diffraction.
\newblock {\em Journal of the Optical Society of America}, 52(2):116--130, 1962.

\bibitem{kitchen2010x}
M.~J. Kitchen, D.~M. Paganin, K.~Uesugi, B.~J. Allison, R.~A. Lewis, S.~B. Hooper, and K.~M. Pavlov.
\newblock X-ray phase, absorption and scatter retrieval using two or more phase contrast images.
\newblock {\em Optics Express}, 18(19):19994--20012, 2010.

\bibitem{kolobov2003fokker}
V.~I. Kolobov.
\newblock {Fokker--Planck modeling of electron kinetics in plasmas and semiconductors}.
\newblock {\em Computational Materials Science}, 28(2):302--320, 2003.

\bibitem{kumar2007direct}
R.~Kumar, S.~K. Kaura, D.~P. Chhachhia, and A.~K. Aggarwal.
\newblock {Direct visualization of Young's boundary diffraction wave}.
\newblock {\em Optics Communications}, 276:54--57, 2007.

\bibitem{leatham2023x}
T.~A. Leatham, D.~M. Paganin, and K.~S. Morgan.
\newblock X-ray dark-field and phase retrieval without optics, via the {Fokker--Planck} equation.
\newblock {\em IEEE Transactions Medical Imaging}, 42(6), 2023.

\bibitem{lim2022low}
H.~Lim, J.~Lee, S.~Lee, H.~Cho, H.~Lee, and D.~Jeon.
\newblock {Low-density foreign body detection in food products using single-shot grid-based dark-field X-ray imaging}.
\newblock {\em Journal of Food Engineering}, 335:111189, 2022.

\bibitem{loo2001new}
B.~W. Loo~Jr, I.~M. Sauerwald, A.~P. Hitchcock, and S.~S. Rothman.
\newblock {A new sample preparation method for biological soft X-ray microscopy: nitrogen-based contrast and radiation tolerance properties of glycol methacrylate-embedded and sectioned tissue}.
\newblock {\em Journal of Microscopy}, 204(1):69--86, 2001.

\bibitem{michelson1995studies}
A.~A. Michelson.
\newblock {\em {Studies in Optics}}.
\newblock University of Chicago Press, Chicago, 1927.

\bibitem{miller2013phase}
E.~A. Miller, T.~A. White, B.~S. McDonald, and A.~Seifert.
\newblock Phase contrast x-ray imaging signatures for security applications.
\newblock {\em IEEE Transactions Nuclear Science}, 60(1):416--422, 2013.

\bibitem{mittal2012making}
A.~Mittal, R.~Soundararajan, and A.~C. Bovik.
\newblock Making a completely blind image quality analyzer.
\newblock {\em IEEE Signal Processing Letters}, 20(3):209--212, 2012.

\bibitem{morgan2020ptychographic}
A.~J. Morgan, H.~M. Quiney, S.~Bajt, and H.~N. Chapman.
\newblock Ptychographic x-ray speckle tracking.
\newblock {\em Journal of Applied Crystallography}, 53(3):760--780, 2020.

\bibitem{Morgan2013}
K.~S. Morgan, P.~Modregger, S.~C. Irvine, S.~Rutishauser, V.~A. Guzenko, M.~Stampanoni, and C.~David.
\newblock A sensitive x-ray phase contrast technique for rapid imaging using a single phase grid analyzer.
\newblock {\em Optics Letters}, 38(22):4605--4608, 2013.

\bibitem{morgan2019applying}
K.~S. Morgan and D.~M. Paganin.
\newblock {Applying the {Fokker--Planck} equation to grating-based x-ray phase and dark-field imaging}.
\newblock {\em Scientific Reports}, 9(1):17465, 2019.

\bibitem{Morgan2011}
K.~S. Morgan, D.~M. Paganin, and K.~K. Siu.
\newblock Quantitative single-exposure x-ray phase contrast imaging using a single attenuation grid.
\newblock {\em Optics express}, 19(20):19781--19789, 2011.

\bibitem{morgan2012x}
K.~S. Morgan, D.~M. Paganin, and K.~K.~W. Siu.
\newblock X-ray phase imaging with a paper analyzer.
\newblock {\em Applied Physics Letters}, 100(12):124102, 2012.

\bibitem{nielsen2017quantitative}
M.~S. Nielsen, K.~B. Damkj{\ae}r, and R.~Feidenhans'l.
\newblock {Quantitative in-situ monitoring of germinating barley seeds using X-ray dark-field radiography}.
\newblock {\em Journal of Food Engineering}, 198:98--104, 2017.

\bibitem{Oksendal2007}
B.~{\O}ksendal.
\newblock {\em {Stochastic Differential Equations: An Introduction with Applications}}.
\newblock Springer-Verlag, Berlin, 6th edition, 2003.

\bibitem{paganin2006coherent}
D.~M. Paganin.
\newblock {\em {Coherent X-ray Optics}}.
\newblock Oxford University Press, 2006.

\bibitem{paganin2019x}
D.~M. Paganin and K.~S. Morgan.
\newblock X-ray {Fokker--Planck} equation for paraxial imaging.
\newblock {\em Scientific Reports}, 9(1):17537, 2019.

\bibitem{paganin2019tutorials}
D.~M. Paganin and D.~Pelliccia.
\newblock X-ray phase-contrast imaging: a broad overview of some fundamentals.
\newblock {\em Advances in Imaging and Electron Physics}, 218:63--158, 2021.

\bibitem{paganin2023paraxial}
D.~M. Paganin, D.~Pelliccia, and K.~S. Morgan.
\newblock Paraxial diffusion-field retrieval.
\newblock {\em Physical Review A}, 108(1):013517, 2023.

\bibitem{pagot2003method}
E.~Pagot, P.~Cloetens, S.~Fiedler, A.~Bravin, P.~Coan, J.~Baruchel, J.~H{\"a}rtwig, and W.~Thomlinson.
\newblock A method to extract quantitative information in analyzer-based x-ray phase contrast imaging.
\newblock {\em Applied Physics Letters}, 82(20):3421--3423, 2003.

\bibitem{pavliotis2014stochastic}
G.~A. Pavliotis.
\newblock {\em {Stochastic Processes and Applications Diffusion Processes, the Fokker--Planck and Langevin Equation}}.
\newblock Springer, 2014.

\bibitem{pavlov2020single}
K.~M. Pavlov, H.~Li, D.~M. Paganin, S.~Berujon, H.~Roug{\'e}-Labriet, and E.~Brun.
\newblock Single-shot x-ray speckle-based imaging of a single-material object.
\newblock {\em Physical Review Applied}, 13(5):054023, 2020.

\bibitem{pavlov2020x}
K.~M. Pavlov, D.~M. Paganin, H.~T. Li, S.~Berujon, H.~Roug{\'e}-Labriet, and E.~Brun.
\newblock X-ray multi-modal intrinsic-speckle-tracking.
\newblock {\em Journal of Optics}, 22(12):125604, 2020.

\bibitem{pavlov2021directional}
K.~M. Pavlov, D.~M. Paganin, K.~S. Morgan, H.~Li, S.~Berujon, L.~Qu{\'e}not, and E.~Brun.
\newblock {Directional dark-field implicit x-ray speckle tracking using an anisotropic-diffusion Fokker--Planck equation}.
\newblock {\em Physical Review A}, 104(5):053505, 2021.

\bibitem{pfeiffer2008hard}
F.~Pfeiffer, M.~Bech, O.~Bunk, P.~Kraft, E.~F. Eikenberry, C.~Br{\"o}nnimann, C.~Gr{\"u}nzweig, and C.~David.
\newblock {Hard-X-ray dark-field imaging using a grating interferometer}.
\newblock {\em Nature Materials}, 7(2):134--137, 2008.

\bibitem{risken1989fokkerplanck}
H.~Risken.
\newblock {\em The {{Fokker--Planck}} {E}quation: {M}ethods of solution and applications}.
\newblock Springer, Berlin, 2nd edition, 1989.

\bibitem{shea1997fokker}
J.-E. Shea and I.~Oppenheim.
\newblock {Fokker--Planck equation and non-linear hydrodynamic equations of a system of several Brownian particles in a non-equilibrium bath}.
\newblock {\em Physica A}, 247(1):417--443, 1997.

\bibitem{shimao2021x}
D.~Shimao, N.~Sunaguchi, T.~Yuasa, M.~Ando, K.~Mori, R.~Gupta, and S.~Ichihara.
\newblock {X-ray Dark-Field Imaging (XDFI)—A Promising Tool for 3D Virtual Histopathology}.
\newblock {\em Molecular Imaging and Biology}, 23:481--494, 2021.

\bibitem{spiegel1959vector}
M.~R. Spiegel.
\newblock {\em {Theory and Problems of Vector Analysis and an Introduction to Tensor Analysis}}.
\newblock Schaum Publishing Company, New York, 1959.

\bibitem{stampanoni2011first}
M.~Stampanoni, Z.~Wang, T.~Th{\"u}ring, C.~David, E.~Roessl, M.~Trippel, R.~A. Kubik-Huch, G.~Singer, M.~K. Hohl, and N.~Hauser.
\newblock The first analysis and clinical evaluation of native breast tissue using differential phase-contrast mammography.
\newblock {\em Investigative Radiology}, 46(12):801--806, 2011.

\bibitem{tan1987solution}
W.~Tan, Y.~Li, and W.~Zhang.
\newblock The solution of the {Fokker--Planck} equation with zero or negative diffusion coefficients in quantum optics.
\newblock {\em Optics Communications}, 64(2):195--199, 1987.

\bibitem{tikhonov1977solutions}
A.~N. Tikhonov and V.~Y. Arsenin.
\newblock {\em {Solutions of Ill-Posed Problems}}.
\newblock Winston, Washington, 1977.

\bibitem{wang2016high}
H.~Wang, Y.~Kashyap, B.~Cai, and K.~Sawhney.
\newblock {High energy X-ray phase and dark-field imaging using a random absorption mask}.
\newblock {\em Scientific Reports}, 6(1):30581, 2016.

\bibitem{wang2016synchrotron}
H.~Wang, Y.~Kashyap, and K.~Sawhney.
\newblock {From synchrotron radiation to lab source: advanced speckle-based X-ray imaging using abrasive pape}r.
\newblock {\em Scientific Reports}, 6(1):20476, 2016.

\bibitem{Wen2010}
H.~H. Wen, E.~E. Bennett, R.~Kopace, A.~F. Stein, and V.~Pai.
\newblock Single-shot x-ray differential phase-contrast and diffraction imaging using two-dimensional transmission gratings.
\newblock {\em Optics Letters}, 35(12):1932--1934, 2010.

\bibitem{wernick2003multiple}
M.~N. Wernick, O.~Wirjadi, D.~Chapman, Z.~Zhong, N.~P. Galatsanos, Y.~Yang, J.~G. Brankov, O.~Oltulu, M.~A. Anastasio, and C.~Muehleman.
\newblock Multiple-image radiography.
\newblock {\em Physics in Medicine \& Biology}, 48(23):3875--3895, 2003.

\bibitem{yashiro2015effects}
W.~Yashiro and A.~Momose.
\newblock Effects of unresolvable edges in grating-based x-ray differential phase imaging.
\newblock {\em Optics Express}, 23(7):9233--9251, 2015.

\bibitem{yashiro2010origin}
W.~Yashiro, Y.~Terui, K.~Kawabata, and A.~Momose.
\newblock {On the origin of visibility contrast in x-ray Talbot interferometry}.
\newblock {\em Optics Express}, 18(16):16890--16901, 2010.

\bibitem{yuen1986langevin}
H.~P. Yuen and P.~Tombesi.
\newblock Langevin equations with negative diffusion coefficients. {A} new approach to quantum optics.
\newblock {\em Optics Communications}, 59(2):155--159, 1986.

\bibitem{zdora2018state}
M.-C. Zdora.
\newblock {State of the art of X-ray speckle-based phase-contrast and dark-field imaging}.
\newblock {\em Journal of Imaging}, 4(5):60, 2018.

\bibitem{zdora2017x}
M.-C. Zdora, P.~Thibault, T.~Zhou, F.~J. Koch, J.~Romell, S.~Sala, A.~Last, C.~Rau, and I.~Zanette.
\newblock X-ray phase-contrast imaging and metrology through unified modulated pattern analysis.
\newblock {\em Physical Review Letters}, 118(20):203903, 2017.

\bibitem{zhang2019particle}
J.~Zhang, B.~John, M.~Pfeiffer, F.~Fei, and D.~Wen.
\newblock Particle-based hybrid and multiscale methods for nonequilibrium gas flows.
\newblock {\em Advances in Aerodynamics}, 1(1):12, 2019.

\end{thebibliography}

\end{document}